\documentclass[fleqn,usenatbib]{mnras}

\usepackage{newtxtext,newtxmath}

\usepackage[T1]{fontenc}

\DeclareRobustCommand{\VAN}[3]{#2}
\let\VANthebibliography\thebibliography
\def\thebibliography{\DeclareRobustCommand{\VAN}[3]{##3}\VANthebibliography}

\usepackage{graphicx}	
\usepackage{amsmath}	
\usepackage{bm}
\usepackage{subcaption}
\usepackage{framed}
\usepackage{aas_macros}
\graphicspath{{figures/}} 

\title[Statistics of eigenvalue fields]{Statistics of tidal and deformation eigenvalue fields in the primordial Gaussian matter distribution: the two-dimensional case}

\author[Feldbrugge, Yan, van de Weygaert]{
Job Feldbrugge,$^{1}$\thanks{E-mail: Job.Feldbrugge@ed.ac.uk}
Yihan Yan,$^{2}$
Rien van de Weygaert$^{3}$
\\
$^{1}$Higgs Centre for Theoretical Physics, University of Edinburgh, James Clerk Maxwell Building, Edinburgh EH9 3FD, United Kingdom\\
$^{2}$DAMTP, University of Cambridge, CB3 0WA Cambridge, United Kingdom\\
$^{3}$Kapteyn Astronomical Institute, University of Groningen, PO Box 800, 9700 AV Groningen, The Netherlands
}

\date{Accepted XXX. Received YYY; in original form ZZZ}

\pubyear{2023}

\begin{document}
\label{firstpage}
\pagerange{\pageref{firstpage}--\pageref{lastpage}}
\maketitle

\begin{abstract}
We study the statistical properties of the eigenvalues of the primordial gravitational tidal and deformation tensor for random Gaussian cosmic density and velocity fluctuation fields. With the tidal and deformation tensors, Hessians of the gravitational and velocity potential, being Gaussian, the corresponding eigenvalue fields are distinctly non-Gaussian. Following the extension of the Doroshkevich formula for the joined distribution of eigenvalues to two-dimensional fields, we evaluate the two- and three-point correlation functions of the eigenvalue fields. In addition, we assess the number densities of singular points of the eigenvalue fields and find their corresponding two- and three-point correlation functions.
  
The incentive for the present study is based on our interest in developing a fully analytical theory for the structure and dynamics of the cosmic web. The role of tidal forces and the resulting mass element deformation in shaping the prominent anisotropic wall-like and filamentary components of the cosmic web has since long been recognized based on the Zel'dovich approximation. Less well-known is that the weblike spatial pattern is already recognizable in the primordial tidal and deformation eigenvalue field, even while the corresponding Gaussian density and the potential field appear merely as a spatially incoherent and unstructured random field. Furthermore, against the background of a full phase-space assessment of structure formation in the Universe, the \textit{caustic skeleton} theory entails a fully analytical framework for the nonlinear evolution of the cosmic web. It accomplishes this by describing the folding characteristics of the dark matter sheet and quantifying the emerging caustic singularities in terms of \textit{caustic conditions}, which are fully specified by the spatial properties of the deformation eigenvalues and eigenvectors. Finally, within the context of tidal torque theory for the generation of the angular momentum of galaxies, tidal tensor eigenvalues are of central importance, and understanding their spatial distribution is a critical element in predicting the resulting rotation amplitude and orientation. 

The current study only applies to two-dimensional Gaussian random fields and will be generalized to a three-dimensional analysis in an upcoming study.
\end{abstract}

\begin{keywords}
    cosmology: theory -- large-scale structure of Universe -- dark matter -- early universe
\end{keywords}



\section{Introduction}
In this paper, we study and analyze the statistics of the primordial tidal and deformation tensor eigenvalue fields. For the dynamical evolution of structure in the Universe, these are of key significance. The emergence, structure, and connectivity of the cosmic web are to be understood in terms of these dynamical quantities. While the primordial matter density fluctuations define with high accuracy, a Gaussian random field, the tidal and deformation eigenvalue fields are distinctly non-Gaussian. Following the seminal contribution by \cite{Doroshkevich:1970}, who derived the joint distribution function of the eigenvalues, we present an extensive analysis of the statistical properties and correlations of eigenvalue fields. 

One of the most significant milestones in modern cosmology is the finding that on Megaparsec scales galaxies, gas and matter are aggregated in a pervasive weblike network. It consists of an intricate connected pattern of \textit{voids}, \textit{walls}, \textit{filaments}, and \textit{clusters} \citep{Zeldovich:1970, Einasto:1977, Bond:1996}. The cosmic web marks a crucial stage in the gravitationally driven evolution of the inhomogeneous cosmic matter distribution from its simple Gaussian primordial conditions to the emergence of intricate nonlinear structures. Predicted by the analytical framework of the Zel'dovich approximation \cite{Zeldovich:1970, Shandarin:1989}, the later detection of the weblike organization of galaxies and gas in the Universe \cite{ Einasto:1978, Lapparent:1986, Colless:2003, Huchra:2012, Granett:2012} confirmed its prominence on scales of a few up to hundreds of Megaparsec. In the coming years a large array of major observational redshift surveys, in particular, those of Euclid, DESI, the Vera Rubin observatory, and SKA -- will map the weblike organization of galaxies over unprecedented large cosmic volumes.

The cosmic web represents a key phase in the dynamical buildup of structure in the Universe. It emerges when the original long phase of linear evolution of the primordial density and velocity field is turning into a more advanced nonlinear stage involving contraction and collapse of mass inhomogeneities. As such, it marks the transition from the primordial (Gaussian) random field to highly nonlinear structures that have fully collapsed into halos and galaxies. In this context, the formation and evolution of anisotropic structures are the product of anisotropic deformations, accurately described by the Zel'dovich formalism in the mildly nonlinear stage \citep{Zeldovich:1970}, and driven by gravitational tidal forces induced by the inhomogeneous mass distribution. Hence, the first recognizable features to emerge are the flattened wall-like and elongated filamentary features, along with the large underdense void regions that assume most of the cosmic volume between these features. Ultimately these merge into a pervasive weblike network. In other words, it is the anisotropy of the force field and the resulting deformation of the matter distribution which are at the heart of the emergence of the weblike structure of the mildly nonlinear mass distribution.

While the seminal role of gravitational tidal force fields in shaping the anisotropic wall-like and filamentary structures in the cosmic web has been recognized for many years \citep{Weygaert:1996, Bond:1996, Weygaert:2008}, an additional major incentive for the present study is the fully nonlinear \textit{caustic skeleton} model of the cosmic web that we have developed in recent years \citep{Arnold:1982b, Hidding:2014, Feldbrugge:2018, Feldbrugge:2022}. It is based on the realization that the evolution of the cosmic web can be understood in detail in terms of the singularities and caustics that are arising in the matter distribution as a result of the structure of the corresponding flow field. It follows one of the most interesting recent developments in our understanding of the dynamical evolution of the cosmic web has been the uncovering of the intimate link between the emerging anisotropic structures and the multistream migration flows involved in the buildup of cosmic structure \citep[][]{Shandarin:2011, Shandarin:2012, Falck:2012, Abel:2012}. The phase-space based \textit{caustic skeleton} description of the evolving weblike pattern in the cosmic matter distribution is centered around a complete set of caustic conditions, which revealed that a full understanding of the cosmic web is obtained through the spatial characteristics of the \textit{eigenvalue} and \textit{eigenvector} fields.  

A substantial body of theoretical and observational evidence underpins the assumption of the Gaussianity of the primordial matter distribution \citep{Adler:1981, Bardeen:1986, Adler:2009} (also see \cite{Pranav:2019}), with only tiny non-Gaussian deviations. The primary evidence for this is the near-perfect Gaussian nature of the Cosmic Microwave Background radiation (CMB) temperature fluctuations. These directly reflect the density and velocity perturbations on the surface of last scattering, and thus the mass distribution at the recombination and decoupling epoch 379,000 years after the Big Bang, at a redshift of $z \approx 1090$ \citep[see e.g.][]{Peebles:1980, Jones:2017}. In particular, the measurements by the COBE, WMAP, and Planck satellites established that to high accuracy the CMB temperature fluctuations define a homogeneous and isotropic Gaussian random field \citep{Smoot:1992, WMAP:2003, Bennett:2003, Spergel:2007, Komatsu:2011, Planck:2020}. A strong second argument for the Gaussian nature and characteristics of these fluctuations is that they narrowly follow the theoretical predictions of the inflationary scenario, at least in its simplest forms \citep{Guth:1981, Linde:1982, Kolb:1990, Liddle:2000}. The inflationary expansion of quantum fluctuations in the generating inflaton (field) leads to a key implication of this process, the generation of cosmic density and velocity fluctuations. It involves the prediction of the resulting density fluctuation field being adiabatic and a homogeneous Gaussian random field, with a near scale-free Harrison-Zel'dovich spectrum \citep{Harrison:1970, Zeldovich:1972, Mukhanov:1981, Guth:1982, Starobinsky:1982, Bardeen:1983}. Third, the {\it Central Limit Theorem} states that the statistical distribution of a sum of many independent and identically distributed random variables will tend to assume a Gaussian distribution. Given that when the Fourier components of a primordial density and velocity field are statistically independent, each having the same Gaussian distribution, then the joint probability of the density evaluated at a finite number of points will be Gaussian \citep{Bardeen:1986}. 

To understand the structure and dynamics of the cosmic web, and to be able to exploit these towards inferring crucial information on cosmology and cosmic structure formation, we need to be able to couple the gravitational evolution process to the primordial conditions out of which the cosmic web arose. While primordial cosmic density and velocity field is a Gaussian random field \citep{Adler:1981, Bardeen:1986, Adler:2009}, fully specified by its power spectrum or, in real space, its correlation function. However, while the density, velocity, and gravitational potential fields are all Gaussian random fields, the situation is distinctly different when turning toward the relevant spatial structure of the tidal and deformation eigenvalue fields. The tidal and deformation fields are the Hessian matrices of the gravitational and velocity potential. While the individual components of these Hessian matrices are Gaussian, the corresponding eigenvalues have a distinctly non-Gaussian character.

To appreciate the intimate relationship between the spatial characteristics of the tidal and deformation eigenvalue field and the emerging cosmic web we refer to figure~\ref{fig:Initial_Condiitions}. In addition to the primordial Gaussian potential and density fields, the top panels, show the distinctively different spatial character of the tidal and deformation eigenvalue field. The bottom panels show the maps of the corresponding eigenvalue fields. The maps reveal that to a considerable extent, a pervasive weblike network can already be recognized in the primordial tidal and deformation field. The maps show distinctly non-Gaussian fields, marking a highly structured pattern, with a high level of spatial coherence. Reflected in the weblike nonlinear matter distribution that emanates out of the primordial spatial pattern of the eigenvalue fields, will provide a substantial increase in insight and understanding of the dynamical evolution of the cosmic web and a transparent path towards connecting the properties of the observed cosmic web with the underlying cosmology.

The spatial structure of the eigenvalue field also contains critical information on the hierarchical buildup and connectivity of structure in the Universe \citep[see][]{Feldbrugge:2018}. This may be immediately understood when assessing the emergence of multistream regions. A local maximum of the eigenvalue field marks the site where we see the appearance of the first nontrivial structures in the matter distribution, the well-known Zel'dovich pancakes. These multi-stream regions grow and connect at the saddle points of the eigenvalue fields to form the web-like structure of the cosmic web. Because of the explicit analytical expressions for this process in the \textit{caustic skeleton} model, in terms of the corresponding caustic condition for the eigenvalue and eigenvector fields, the statistics of the eigenvalue field will enable us to obtain a detailed analytical and statistical inventory at different redshifts of a large range of properties of the cosmic web and its various constituents.

The spatial properties of the tidal eigenvalue fields are also of critical importance in understanding the rotation of galaxies. The same tidal forces that shape the cosmic web are known to be the source of the angular momentum of contracting and collapsing matter halos and the galaxies they contain. Thus, the same tidal fields that drive the formation of structures in the cosmic web also result in the acquisition of angular momentum of proto-halos taking shape in these structures. According to Tidal Torque Theory \citep{Hoyle:1951, Peebles:1969, Doroshkevich:1970, White:1984} the tidal fields exert a torque, inducing the rotation of a contracting protohalo as a result of the differential orientation between its inertia tensor and the local gravitational tidal tensor. Accordingly, the theory suggests a direct correlation between halo properties such as angular momentum, and the shape and the orientation of host structures \citep{Porciani:2002a, Porciani:2002b, Schaffer:2009}. Several studies have attempted to connect the resulting (nonlinear) angular momentum to the spatial structure of the primordial matter distribution \citep{Lee:2000, Cadiou:2022}, and further progress will substantially profit from insight into the tidal eigenvalue statistics. 

In the current paper, we assess the statistics of primordial (and Lagrangian) tidal and deformation tensor eigenvalue fields. This concerns the evaluation of the one-point functions of the eigenvalue fields, \textit{i.e.}, their PDF, as well as the characterization of their spatial structure in terms of two-point correlation functions, as well as three-point and higher order functions. It extends the well-known statistical properties of Gaussian random fields \citep{Adler:1981, Bardeen:1986, Adler:2009}  (also see \cite{Park:2013, Pranav:2019, Feldbrugge:2019} for the topological properties of Gaussian random fields) to fields that have a distinctly non-Gaussian character. It involves the explicit case of the eigenvalue fields of Gaussian tidal and deformation tensor fields, themselves the Hessians of the Gaussian gravitational or deformation potential fields. Recognizing the instrumental significance of (primordial) tidal fields in the structure formation process \citep[see][for extensive discussions]{BondMyers:1996, Weygaert:2008}, various statistical aspects of these fields have been discussed. The pioneering study of Doroshkevich \citep{Doroshkevich:1970} inferred the PDF of the combined eigenvalue fields, instrumental in assessing the expected distribution of fully collapsed mass concentrations, expanding void regions, and wall-like and filamentary structures in the mass distribution, as foreseen by the Zel'dovich approximation \citep{Zeldovich:1970}. The Gaussian one-point statistics of the components of the tidal and deformation tensor by \cite{Weygaert:1996}, was elaborated upon in an analytical study of the spatial correlations of the tidal tensor components by \cite{Catelan:2001}. Later studies turned to the more complex -- and physically highly relevant -- aspect of the statistical properties of the corresponding eigenvalue fields \citep{Lee:1998, Desjacques:2008, Rossi:2012}. \cite{Lee:1998} concentrated on the one-point conditional distribution function of eigenvalues, along with an estimate of the implied clump mass functions. Also, \cite{Rossi:2012} concentrated on the one-point distribution of tidal eigenvalues, focussing on that around peaks and dips in the density field. The issue of the implied spatial structure of the tidal eigenvalue fields was addressed by \cite{Desjacques:2008}, in terms of their two-point correlation function. The treatment limited itself to various asymptotic limits, which were assumed to be near Gaussian. As we need the full scope of the non-Gaussian eigenvalue statistics to be enabled to describe and analyze the distinctly non-Gaussian pattern of the cosmic web, and to analyze it in terms of the caustic skeleton model \citep{Feldbrugge:2018}, the present study contains a complete statistical treatment of the non-Gaussian tidal and deformation eigenvalue fields. This involves the one-point distribution functions, \textit{i.e.}, the PDFs, as well as their spatial characterization in terms of two-point and three-point correlation functions. Particularly interesting for understanding the complex connectivity of the cosmic web is the statistical analysis of the singularities -- maxima, minima, and saddle points -- in the eigenvalue fields. In terms of the dynamical evolution of the cosmic web, it is these points that determine the emergence of the various structural features of the cosmic web and their merging and assembly into larger weblike complexes. 

In this paper, we first give a concise summary of Gaussian random field theory in section \ref{sec:GRF}. Following definitions and preliminaries concerning the eigenvalue fields of tidal and deformation tensors in section~\ref{sec:eigenvalue_prelim}, we study the spatial statistical properties of the eigenvalue fields in section~\ref{sec:eigenvalue_stat}. In this section, we focus on the PDF, and the two-point and three-point correlation functions of these fields. It involves, amongst others, the extension of the Doroshkevich formula for the PDF of three-dimensional eigenvalue fields to two-dimensional fields. Next, in section \ref{sec:singular}, we investigate the number densities of singular points of the eigenvalue fields and find their corresponding two- and three-point correlation functions. We summarize the results in section \ref{sec:Conclusion}. The current study only applies to two-dimensional Gaussian random fields. However, the techniques generalize to the three-dimensional case, which will be addressed in a follow-up study.

\section{Gaussian random field}\label{sec:GRF}
Gaussian random fields occur commonly in nature. Examples can be found in the random noise in telephone lines, the height maps of ocean waves and mountain ranges, and the statistical fluctuations in the density field at the epoch or recombination. In many circumstances, Gaussian random fields form as the random superposition of features, following the central limit theorem \citep{Feynman:1965}. The vacuum fluctuations of a free quantum field theory are another example of a Gaussian random field. In cosmology, it is often assumed that the density fluctuations at the time or recombination are a remnant of these quantum fluctuations in the early universe. 

A two-dimensional Gaussian random field $f:\mathbb{R}^2\to \mathbb{R}$ is a generalization of a multi-dimensional normal distribution to the continuum, defined by the probability density
\begin{align}
p(f) = \mathcal{N} e^{- S[f]}\,, \label{eq:functional_Distribution}
\end{align}
with the normalization constant $\mathcal{N}$ and the `action' (in analogy with the Euclidean path integral \citep[see][]{Feynman:1965})
\begin{align}
S[f]\equiv \frac{1}{2} \iint [f(\bm{q}_1) - \bar{f}(\bm{q}_1)] K(\bm{q}_1,\bm{q}_2) [f(\bm{q}_2) -\bar{f}(\bm{q}_2)]\mathrm{d}\bm{q}_1 \mathrm{d}\bm{q}_2,\label{eq:action}
\end{align}
defined in terms of the mean-field $\bar{f}(\bm{q})$ and the kernel $K(\bm{q}_1,\bm{q}_2)$ \citep{Longuet-Higgins:1957, Adler:1981, Bardeen:1986, Adler:2009}. The probability that the random field $f$ is included in a set of functions $\mathcal{S}$ is defined by the path integral
\begin{align}
P[f \in \mathcal{S}] = \mathcal{N} \int \bm{1}_\mathcal{S}(f) e^{-S[f]}\,\mathcal{D}f\,,
\end{align}
with $\bm{1}_\mathcal{S}$ the identity function\footnote{Defined by $\bm{1}_\mathcal{S}(x)=1$ when $x \in \mathcal{S}$ and $\bm{1}_\mathcal{S}(x)=0$ when $x \notin \mathcal{S}$.} and $\mathcal{D}f$ the path integral measure. The expectation value of a functional $Q[f]$ is given by
\begin{align}
\left\langle Q[f] \right\rangle = \mathcal{N}\int Q[f]\, e^{-S[f]}\,\mathcal{D}f\,,
\end{align}
analogous to the Euclidean path integrals in statistical field theory. It can be shown that the expectation value of the Gaussian random field is given by the mean-field
\begin{align}
\langle f(\bm{q})\rangle &= \bar{f}(\bm{q})\,,
\end{align}
and that the two-point correlation function
\begin{align}
\xi(\bm{q}_1, \bm{q}_2) &= \langle (f(\bm{q}_1) - \bar{f}(\bm{q}_1)) (f(\bm{q}_2) - \bar{f}(\bm{q}_2))\rangle\nonumber\\
&= \int (f(\bm{q}_1) - \bar{f}(\bm{q}_1)) (f(\bm{q}_2) - \bar{f}(\bm{q}_2)) e^{-S[f]}\mathcal{D}f
\end{align}
is the inverse of the kernel $K$, \textit{i.e.,}
\begin{align}
\int K(\bm{q}_1,\bm{q}) \xi(\bm{q},\bm{q}_2) \mathrm{d}\bm{q}= \delta_D^{(2)}(\bm{q}_1-\bm{q}_2)\,,\label{eq:defK}
\end{align}
with the two-dimensional Dirac delta function $\delta_D^{(2)}$. The Gaussian random field is thus fully determined by the mean-field $\bar{f}$ and the two-point correlation function $\xi$. 

In cosmology, the cosmological principle often leads to the study of statistically homogeneous and isotropic random fields for which the mean field is constant $\bar{f}(\bm{q})=\bar{f}=0$ and the two-point correlation function only depends on the magnitude of the difference of the inserted points, \textit{i.e.}, $\xi(\bm{q}_1,\bm{q}_2)=\xi(\bm{q}_1-\bm{q}_2) = \xi(\|\bm{q}_1-\bm{q}_2\|)$, and consequently $K(\bm{q}_1,\bm{q}_2)=K(\bm{q}_1-\bm{q}_2)=K(\|\bm{q}_1-\bm{q}_2\|)$. 

The statistical properties of homogeneous and isotropic random fields are most transparently expressed in terms of the Fourier transform of the random field
\begin{align}
\hat{f}(\bm{k}) = \int f(\bm{q})e^{i\bm{k}\cdot \bm{q}}\mathrm{d}\bm{q}\,,
\end{align}
satisfying the reality condition $\hat{f}(\bm{k})=\hat{f}^*(-\bm{k})$, with the inverse Fourier transform
\begin{align}
f(\bm{q}) = \int \hat{f}(\bm{k})e^{-i \bm{k} \cdot \bm{q}}\frac{\mathrm{d}\bm{k}}{(2\pi)^2}\,.
\end{align}
Using the double convolution theorem, we express the action \eqref{eq:action} as the single integral
\begin{align}
S[f] = \frac{1}{2} \int |\hat{f}(\bm{k})|^2 \hat{K}(\bm{k}) \frac{\mathrm{d}\bm{k}}{(2\pi)^2}\,.
\end{align}
In Fourier space, equation \eqref{eq:defK} takes the form
\begin{align}
\int \hat{K}(\bm{k})\, P(\bm{k})\, e^{i\bm{k}(\bm{q}_1-\bm{q}_2)} \frac{\mathrm{d}\bm{k}}{(2\pi)^2} = \delta_D^{(2)}(\bm{q}_1 - \bm{q}_2)\,,
\end{align}
with the power spectrum defined as the Fourier transform of the two-point correlation function,
\begin{align}
P(\bm{k}) = \int \xi(\bm{q}) \, e^{i\bm{k}\cdot \bm{q}}\mathrm{d}\bm{q}\,,
\end{align}
implying the relation $\hat{K}(\bm{k}) = 1/P(\bm{k})$. The resulting probability density is diagonal in the Fourier modes
\begin{align}
p(\hat{f}) \propto \exp\left[ -\frac{1}{2} \int \frac{|\hat{f}(\bm{k})|^2}{P(\bm{k})} \frac{\mathrm{d}\bm{k}}{(2\pi)^2}\right]\,,
\end{align}
implying the covariance of the Fourier modes
\begin{align}
\langle \hat{f}(\bm{k}_1)\hat{f}^*(\bm{k}_2) \rangle = (2\pi)^2 \delta_D^{(2)}(\bm{k}_1-\bm{k}_2) P(\bm{k}_1)\,.
\end{align}

In practice, we often consider realizations of Gaussian random fields on a lattice, or more generally a finite set of linear statistics of the random field $\bm{Y}$, consisting of the random field, a derivative in a point or more generally a convolution of the random field. In this setting, the functional distribution \eqref{eq:functional_Distribution} reduces to the multi-dimensional Gaussian distribution,
\begin{align}
p(\bm{Y}) = \frac{\exp\left[-\frac{1}{2}  \Delta \bm{Y}^T M^{-1} \Delta \bm{Y}\right]}{[(2\pi)^n \det M]^{1/2}}\,,
\end{align}
with the length $n$ of the vector $\bm{Y}$, the deviation from the mean $\Delta \bm{Y} = \bm{Y} - \langle \bm{Y}\rangle$ and the covariance matrix
\begin{align}
M = \text{cov}(\bm{Y},\bm{Y}) = \langle \Delta \bm{Y}^T \Delta \bm{Y}\rangle\,.
\end{align}
When generating a random field in Fourier space, consider the distribution of the discrete Fourier modes $\bm{Y}=(\hat{f}(\bm{k}_1),\hat{f}(\bm{k}_2),\dots)$,
\begin{align}
p\left(\hat{f}(\bm{k}_1), \hat{f}(\bm{k}_2), \dots\right) = \prod_{i} \frac{1}{\sqrt{2\pi P( \bm{k}_i)}} \exp\left[-\frac{|\hat{f}(\bm{k}_i)|^2}{2P(\bm{k}_i)}\right]\,.
\end{align}
The Fourier modes are independently and normally distributed with the variance $P(\bm{k}_i)$.

The statistical properties of random fields are often conveniently expressed in terms of the moments
\begin{align}
\sigma_i^2 &= \frac{1}{(2\pi)^2} \int \|\bm{k}\|^{2i}P(\bm{k})\mathrm{d}\bm{k}\nonumber\\
&= \frac{1}{2\pi} \int_0^\infty k^{2i+1}P(k)\mathrm{d}k\,,
\end{align}
with the magnitude $k = \|\bm{k}\|$. The first moments $\sigma_0^2,\sigma_1^2$, and $\sigma_2^2$ can be interpreted as the variance $\langle f^2 \rangle$, the variance of the norm of the gradient $\langle \|\nabla f\|^2\rangle$ and the variance of the Laplacian $\langle (\nabla^2 f)^2\rangle$ of the random field. In terms of the two-point correlation function, we find the relation $\sigma_i^2=(-\nabla^{2})^i\xi(\bm{0})$.

\section{Eigenvalue fields: preliminaries}
\label{sec:eigenvalue_prelim}
In section~\ref{sec:GRF}, we defined the Gaussian random field as a random process that is completely characterized by the mean and two-point correlation functions. The primordial density field and corresponding gravitational potential are examples of Gaussian random fields. The tidal gravitational field is the Hessian of the gravitational potential, and the corresponding deformation field in the Zel'dovich approximation is the Hessian of the Lagrangian velocity potential. Each of the $3 \times 3$ individual components of these Hessian tensors also defines Gaussian random fields. However, the eigenvalues of the Hessian tensor represent fields that are distinctly non-Gaussian. The eigenvalue fields are non-Gaussian fields with richer geometry and non-trivial higher-order correlation functions. Caustic skeleton theory suggests that these non-Gaussianities might carry over into the non-Gaussian nature of the present-day
cosmic web \citep{Feldbrugge:2022}. 

In the present section, we focus on the eigenvalue fields of the primordial Gaussian tidal and deformation tensor. 

\begin{figure*}
    \centering
    \begin{subfigure}[b]{0.49\textwidth}
        \includegraphics[width=\textwidth]{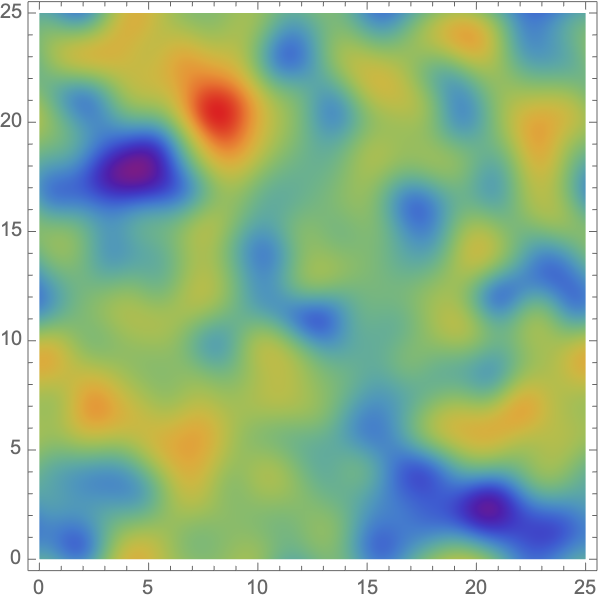}
        \caption{$\phi$}\label{fig:Phi}
    \end{subfigure}~
    \begin{subfigure}[b]{0.49\textwidth}
        \includegraphics[width=\textwidth]{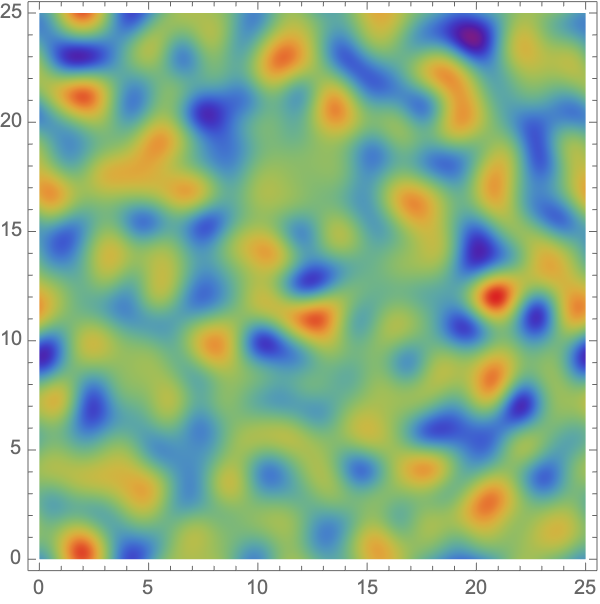}
        \caption{$\delta$}\label{fig:delta}
    \end{subfigure}\\
    \begin{subfigure}[b]{0.49\textwidth}
        \includegraphics[width=\textwidth]{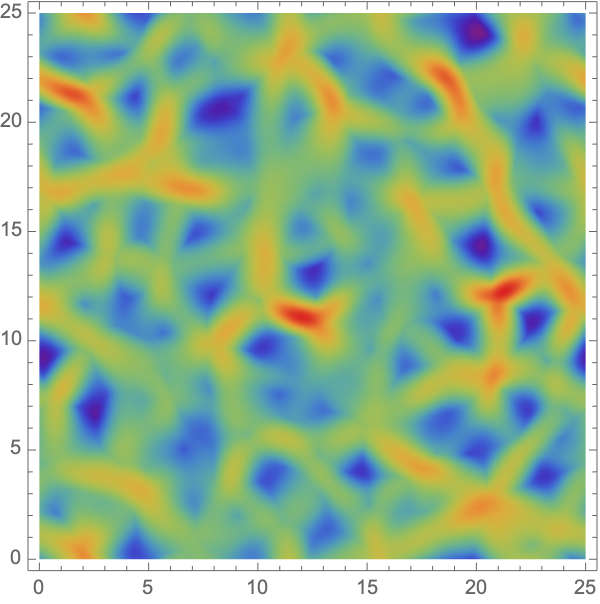}
        \caption{$\lambda_1$}\label{fig:lambda_1}
    \end{subfigure}~
    \begin{subfigure}[b]{0.49\textwidth}
        \includegraphics[width=\textwidth]{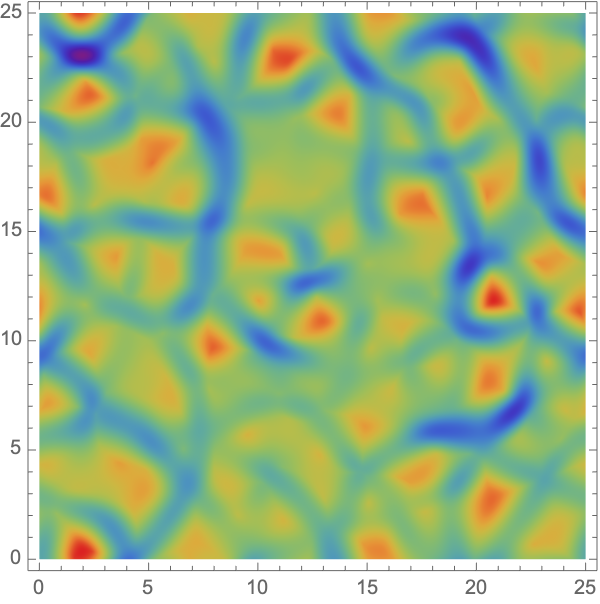}
        \caption{$\lambda_2$}\label{fig:lambda_2}
    \end{subfigure}
    \caption{The gravitational potential (upper left), the density perturbation (upper right), and the two corresponding eigenvalue fields (lower left and lower right).}
    \label{fig:Initial_Condiitions}
\end{figure*}

\subsection{Tidal and Deformation tensors}\label{sec:prelim}
In cosmology, we often work in terms of the primordial density perturbation
\begin{align}
  \delta = (\rho - \bar{\rho})/\bar{\rho}\,.
\end{align}
The primordial density field is the realization of a Gaussian random field. The corresponding potential field $\phi$ is linearly related to the density field through the Poisson equation
\begin{align}
\nabla^2\phi = 4 \pi G \bar{\rho} a^2 \delta \,,
\end{align}
with Newton's gravity constant $G$ and the scale factor $a$.

The tidal force $E_{ij}$ exerted on a mass element is the differential gravitational force, and hence the Hessian
of the gravitational potential $\phi$. In Lagrangian space, it is defined as
\begin{align}
   {\bm E} ({\bm q}) = \begin{pmatrix} E_{11}(\bm{q}) & E_{12}(\bm{q}) \\ E_{12}(\bm{q}) & E_{22}(\bm{q})\end{pmatrix}\ ,
\end{align}
with $E_{ij}({\bm q})  = {\displaystyle \partial^2 \phi_0({\bm q}) \over \displaystyle \partial q_i \partial q_j}$,
in which $\phi_0$ is the primordial gravitational potential linearly extrapolated to the current epoch.

Note that often the tidal field tensor is taken to be the traceless form by subtracting a term proportional to the density $\delta$ from the diagonal components. The physically relevant amplitudes of the tidal tensor are its eigenvalues $T_1$ and $T_2$, expressing its strength along the main direction of the tidal ellipsoids, \textit{i.e.}, along the directions specified by the corresponding eigenvectors $\bm{v}_{t,1}$ and $\bm{v}_{t,2}$,
  \begin{align}
    \bm{E} \bm{v}_{t,i} = E_i \bm{v}_{t,i}\,.
\end{align}
Rotating to the eigenframe, the tidal tensor turns into the diagonal matrix,
\begin{align}
  {\bm E}(\bm{q}) \,=\, \begin{pmatrix} E_{1}(\bm{q}) & 0 \\ 0 & E_{2}(\bm{q})\end{pmatrix}.
\end{align}

\bigskip
Closely related to the primordial tidal tensor $E_{ij}({\bm q})$ is the induced deformation of mass elements. In Lagrangian fluid dynamics, the cosmic matter distribution is described in terms of mass elements, which expand, contract, and twist while conserving their mass. At time $t$, a mass element starting from the position $\bm{q}$ has moved to the position
\begin{align}
    \bm{x}_t(\bm{q}) = \bm{q} + \bm{s}_t(\bm{q})\,,
\end{align}
The gradient $\nabla_{\bm q} \bm{s}_t$ of the displacement describes the deformation of the moving mass element. In case the motion of the mass element is a potential flow, the displacement $\bm{s}_t(\bm{q})$ is the gradient of a potential field. In the cosmological context, this concerns the linearly growing displacement potential $\Psi$, 
\begin{align}
  \bm{s}_t(\bm{q}) = -b_+(t) \nabla \Psi(\bm{q})\,.
\end{align}  
in which the linear growth of the potential includes the (linear) growing mode term $b_+$ and Lagrangian displacement potential $\Psi({\bm q})$. The corresponding deformation of the mass element, quantified in terms of the deformation tensor $\psi$, is the Hessian of the displacement potential,
\begin{align}
    \bm{\psi} = \left[ \frac{\partial^2 \Psi}{\partial q_i \partial q_j} \right]_{i,j=1,2}\,.
\end{align}
The eigenvalues $\lambda_i$ of the deformation tensor $\bm{\psi}$, are defined by the eigen equation
\begin{align}
    \bm{\psi} \bm{v}_i = \lambda_i \bm{v}_i\,,
\end{align}
with the eigenvectors $\bm{v}_i$ and the ordered eigenvalues $\lambda_1 \geq \lambda_2$.

\bigskip
In the cosmological context, at the initial linear phase of structure evolution, there is $1$-$1$ linear relation between the tidal tensor $E_{ij}$ and the deformation tensor $\psi_{ij}$. According to the first-order Zel'dovich approximation (see sect.~\ref{sec:Zeldovich}), 
\begin{align}
  \psi_{ij}\,&\equiv \frac{1}{4 \pi G {\bar \rho} a^2} \frac{\partial^2 \phi}{\partial q_i \partial q_j}\\
  \,&=\,\frac{2 }{3\Omega_0 H_0^2}\,{\displaystyle \partial^2 \phi_0({\bm q}) \over \displaystyle \partial q_i \partial q_j}\equiv \frac{\partial^2 \phi}{\partial q_i \partial q_j}\\
  \,&=\,\frac{2 }{3\Omega_0 H_0^2}\,E_{ij}\,.
  \end{align}

\subsection{Eigenvalue Fields \& Primordial Cosmic Web:\\ \ \ \ \ \ \ \ \ \ illustration}
For a visual appreciation of the different structural character of the Gaussian density and potential field on the one hand and the distinctly non-Gaussian eigenvalue fields, figure~\ref{fig:Initial_Condiitions} shows a realization of a Gaussian random density field and the corresponding potential and deformation eigenvalue fields. 

The primordial density field $\rho$, shown in figure~\ref{fig:delta}, is the realization of a Gaussian random field with the mean density $\bar{\rho}$. In cosmology, we usually work in terms of the primordial density perturbation
\begin{align}
  \delta = \frac{(\rho - \bar{\rho})}{\bar{\rho}}\,,
\end{align}
which is also a realization of a Gaussian random field following the same power spectrum and a vanishing mean. For our case study, we assume a power spectrum
\begin{align}
    P_\delta(k) \propto k^{n_s}e^{-R_s^2 k^2}\,,
\end{align}
with spectral index $n_s$, corresponding to a primordial scale-free field, and Gaussian cutoff scale $R_s$. For the illustrations in this paper, we consider a scale-free spectral index $n_s=3$ and a unit cutoff $R_s=1$. Noteworthy is the statistical homogeneity and isotropy of the density field realization in figure~\ref{fig:delta}. As far as the spatial structure is concerned, it is fully specified by the corresponding two-point correlation function, 
\begin{align}
    \xi_\delta(r) \propto L_{-1-n_s/2}\left(-\frac{r^2}{4R_s^2}\right)\,,
\end{align}
with the Laguerre polynomial $L_n$. For a Gaussian field, the two-point function $\xi(r)$ fully specifies the spatial structure of the field. Moreover, it is important to see that $\xi(r)$ is isotropic, only dependent on radial distance $r$, and therefore has a rather limited scope towards the structural complexity it may entail. 

\bigskip
Also, the corresponding gravitational potential field $\phi$ is a statistically homogeneous and isotropic Gaussian field. This may be directly inferred from the linear relationship between gravitational potential and density field given by the Poisson equation,
\begin{align}
  \nabla^2\phi = 4 \pi G \bar{\rho} a^2 \delta\,,
  \end{align}
with Newton's gravity constant $G$ and scale factor $a$. The corresponding potential power spectrum is 
\begin{align}
    P_\phi(k) \propto k^{n_s-4}e^{-R_s^2 k^2}\,,
\end{align}
with the corresponding two-point correlation function
\begin{align}
    \xi_\phi(r) \propto L_{1-n_s/2}\left(-\frac{r^2}{4R_s^2}\right)\,.
\end{align}
By the Poisson equation, the two-point correlation functions of the density perturbation and gravitational potential are related by the identity $\xi_\delta \propto \nabla^4 \xi_\phi$. The corresponding realization of the potential field is illustrated in figure~\ref{fig:Phi}.  

\bigskip
Turning to the corresponding deformation tensor eigenvalue fields, shown in figures~\ref{fig:lambda_1} and ~\ref{fig:lambda_2}, we observe structural patterns that provide a telling contrast to that of the density and potential fields.

While the eigenvalue fields are derived from a Gaussian random field, they are themselves not Gaussian (when the dimension of the random field is larger than one). Specifically, for the two-dimensional case, the eigenvalues are non-linearly related to the gravitational potential as they are the roots of the quadratic characteristic polynomial $\det(M - \lambda I)$.  In three dimensions, the eigenvalues are cubic roots, potentially leading to stronger non-Gaussian features. The eigenvalue fields are nonetheless closely related to Gaussian ones. By the Poisson equation, the sum over the eigenvalue fields is Gaussian, \textit{i.e.},
\begin{align}
    \lambda_1 + \lambda_2 
    =\nabla^2 \Psi
    = \frac{2}{3\Omega_0 H_0^2} \nabla^2 \phi_0 
    = \frac{2}{3 b_+ a^2  H^2 \Omega} \nabla^2 \phi_{\text{linear}}
    =\delta\,.\label{eq:lambda_delta}
\end{align}

The non-Gaussian nature of the eigenvalue fields is most directly seen when observing the elongated filamentary structures in figures \ref{fig:lambda_1} and \ref{fig:lambda_2}. This is in particular so when noting that the elongated line-like features in the first eigenvalue field $\lambda_1$ are associated with the first collapse in the Zel'dovich approximation. They are the progenitors of the filaments of the cosmic web (in the three-dimensional context, they would be the walls of the cosmic web). Where these elongated filaments meet at nodal joints of the weblike pattern, we see the formation of a cluster.

Motivated by caustic skeleton theory \citep{Feldbrugge:2018, Feldbrugge:2022}, we argue that the association of singular points of the deformation tensor to the geometric features of the cosmic web is more intricate and complex than the conventional assumption relating features of the cosmic web to critical points of the (primordial) density field. This insight is based on the fact that it involves a deeper understanding of the dynamics of gravitational structure formation, involving a full phase-space assessment of the process. Within this context, the local minima of the first eigenvalue field mark the locations that become emptier over time: to first-order approximation they correspond to the cosmic voids of the cosmic web. Meanwhile, the saddle points in the eigenvalue field relate to the walls and filaments of the weblike network, while the local maxima correspond to clusters at the nodes of the cosmic web. Caustic skeleton theory reveals the existence of a larger pallet of singularities, in particular, that of umbilic points that form -- in the two-dimensional context -- an additional set of clusters \footnote{in the three-dimensional context they define one of the two classes of filaments}.

In summary, the eigenvalue fields reveal a substantially richer geometry, reflected also in non-trivial higher-order correlation functions. The impression is that of the embryonic form of the cosmic web that will emerge as a result of the gravitationally driven evolution in the subsequent billions of years of cosmic evolution. Indeed, this is exactly what the caustic skeleton theory of cosmic web formation stipulates \citep{Feldbrugge:2018, Feldbrugge:2022}: it establishes the central role of deformation eigenvalues and eigenvectors in the emergence of the structural pattern of the cosmic web, and hence how the primordial non-Gaussianities of these fields carry over into the non-Gaussian nature of the present-day cosmic web.

\subsection{Zel'dovich approximation}
\label{sec:Zeldovich}
The 1-1 relation between tidal tensor and deformation tensor in the linear regime is directly following from the Zel'dovich approximation. Describing structure formation in terms of Lagrangian perturbation theory it is based on the first-order approximation of the displacement $\bm{s}_t(\bm{q})$ of mass elements at primordial location $\bm{q}$,
\begin{align}
  \bm{x}_t(\bm{q}) = \bm{q} + \bm{s}_t(\bm{q})\,,
\end{align}
with
\begin{align}
  \bm{s}_t(\bm{q}) = -b_+(t) \nabla \Psi(\bm{q})\,,
\end{align}
in which $b_+$ the linear growing mode and $\Psi$ the displacement potential. The latter, proportional to the gravitational potential $\phi_0$ linearly extrapolated to the current epoch $t_0$,
\begin{align}
    \Psi(\bm{q}) 
    = \frac{\phi}{4 \pi G \bar{\rho} a^2} 
    = \frac{2 \phi_0(\bm{q})}{3\Omega_0 H_0^2} \,,
\end{align}
encapsulates the spatial structure of the induced mass streams.  Hence, according to the Zel'dovich approximation, the mass elements follow ballistic trajectories with $b_+$ functioning as the parameterization of time. The latter, which may be considered as the natural time parameter for cosmic structure formation, is the solution of the 2nd-order differential equation
\begin{align}
    \frac{\mathrm{d}^2 b_+(t)}{\mathrm{d}t^2} + 2 \frac{\dot{a}(t)}{a(t)} \frac{\mathrm{d}b_+(t)}{\mathrm{d}t} = 4 \pi G \bar{\rho}(t) b_+(t)\,,
\end{align}
with the boundary condition $b_+(t_0)=1$. In the above, $H_0$ is the current Hubble parameter, and $\Omega_0=\bar{\rho}_0/\rho_c$ the mean matter density in terms of the critical density \footnote{In general, $\Omega = 8 \pi G \bar{\rho}/(3H^2)$ with the Hubble parameter $H$, Newton's constant $G$, and the mean density $\bar{\rho}$.}.

\bigskip
The Zel'dovich approximation \citep{Zeldovich:1970} entails a remarkably accurate specification of the evolving mass distribution up to the phase at which mass streams cross and multistream regions emerge. As such, the Zel'dovich approximation led to the prediction of the formation of anisotropic planar and elongated structures marking the cosmic matter distribution and hence of the cosmic web \citep[see][]{Shandarin:2009}, many years before the observational evidence and confirmation for its existence. Equally important is the central role the formalism fulfills in a large number of theoretical and numerical developments towards understanding structure formation, and instruments for the analysis of the observed Megaparsec scale galaxy and mass distribution.

\begin{figure*}
    \centering
    \begin{subfigure}[b]{0.48\textwidth}
        \includegraphics[width=\linewidth]{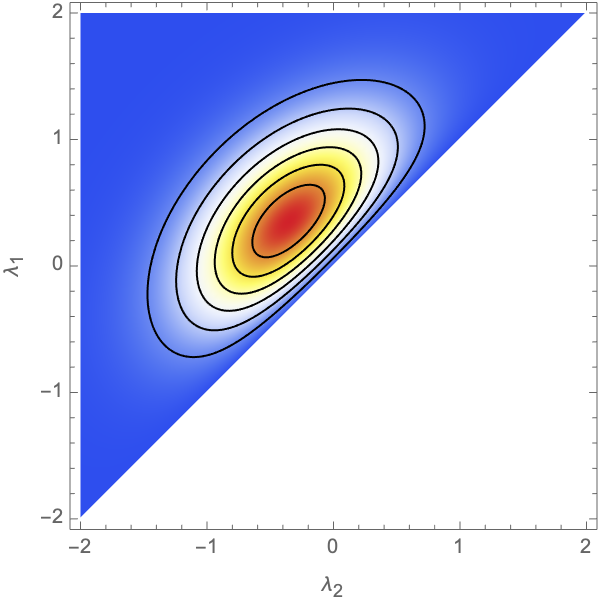}
        \caption{The joined PDF of the eigenvalue fields $p(\lambda_1,\lambda_2)$ is given by the Doroshkevich formula in units of $\sigma_\delta$.}\label{fig:Doroshkevich}
    \end{subfigure}\quad \quad
    \begin{subfigure}[b]{0.48\textwidth}
        \includegraphics[width=\linewidth]{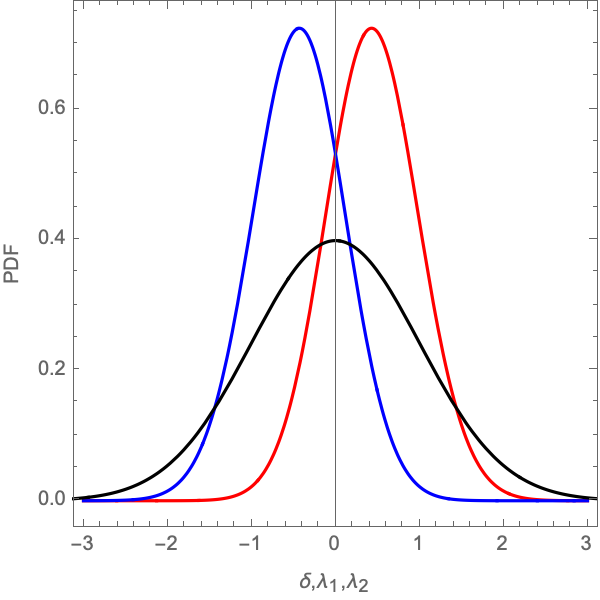}
        \caption{The PDF of the eigenvalue fields $p(\lambda_1)$ (red),$p(\lambda_2)$ (blue) and the density perturbation $p(\delta)$ in units of $\sigma_\delta$.}\label{fig:PDF}
    \end{subfigure}
    \caption{The joined and marginal PDFs of the eigenvalue fields compared to the PDF of the primordial density perturbation field.}\label{fig:PDFs}
\end{figure*}    

Of key significance is the density evolution implied by the Zel'dovich approximation. Using the deformation tensor,
\begin{align}
  \bm{\psi}\,=\, - \nabla \bm{x}_t(\bm{q})\,,
\end{align}
it takes into account the changing volume of mass elements as they expand, contract, and change shape while conserving mass. The resulting density of a Lagrangian mass element follows from the reciprocal of the deformation tensor,
\begin{align}
    \rho(\bm{x}) = \sum_{\bm{q} \in \bm{x}_t^{-1}(\bm{q})} \frac{\bar{\rho}}{|\det \nabla \bm{x}_t(\bm{q})|}\,.\label{eq:densdens}
\end{align}
The deformation tensor $\psi$ is the Hessian of the deformation potential $\Psi$,
\begin{align}
    \bm{\psi} \,=\,\left[ \frac{\partial^2 \Psi}{\partial q_i \partial q_j} \right]_{i,j=1,2}\,,
\end{align}
and as such is proportional to the Hessian of the primordial gravitational potential, \textit{i.e.}, to the primordial tidal field $E_{ij}$ (see sect.~\ref{sec:prelim}). Evaluating the determinant of the deformation tensor $\bm{\psi}$ in terms of its eigenvalues $\lambda_i$,
\begin{align}
    \bm{\psi} \bm{v}_i = \lambda_i \bm{v}_i\,,
\end{align}
in which $\bm{v}_i$ are the corresponding eigenvectors, we obtain the expression for the implied density evolution,
\begin{align}
    \rho(\bm{x}) = \sum_{\bm{q} \in \bm{x}_t^{-1}(\bm{q})} \frac{\bar{\rho}}{|1-b_+(t) \lambda_1(\bm{q})||1-b_+(t) \lambda_2(\bm{q})|}\,.
\end{align}

\bigskip
\noindent Hence, the Zel'dovich formalism implies that a mass element undergoes shell-crossing at a time $t$, at which 
\begin{align}
    1-b_+(t)\lambda_i(\bm{q})=0\,.
\end{align}

\section{Eigenvalue fields: PDF and spatial correlations}\label{sec:eigenvalue_stat}
Following the preliminaries and definitions in the previous section, we are set to assess the statistical properties of the eigenvalue fields. First, we consider the PDF of the eigenvalue fields, followed by the evaluation of two-point and three-point correlation functions.

To this end, we note that from hereon we work from the perspective of the displacement potential. Unless otherwise mentioned, the power spectrum will denote the spectrum of the displacement potential. Sometimes, we will drop the corresponding subscripts, \textit{i.e.}, $P=P_\Psi$ and $\xi=\xi_\Psi$. In this context, $\sigma_2^2$ is the variance of the density perturbation $\sigma_\delta^2$. The results obtained in this paper are expressed in units of this standard deviation $\sigma_\delta$ and the Gaussian spectral cutoff scale $R_s$.

\subsection{The Doroshkevich formula}
The joined PDF of the eigenvalue fields of the three-dimensional Gaussian random field is given by the Doroshkevich formula \citep{Doroshkevich:1970}. We here derive the two-dimensional analog. For convenience, we write the partial derivative of the gravitational potential with a shorthand notation, \textit{i.e.}, $T_{i_1\dots i_n} = \partial^n \Psi/\partial q_{i_1} \dots \partial q_{i_n}$, yielding a concise form of the deformation tensor in the Zel'dovich approximation
\begin{align}
    \bm{\psi} = \begin{pmatrix}
        T_{11} & T_{12}\\
        T_{12} & T_{22}
    \end{pmatrix},
\end{align}
and the eigenvalue fields
\begin{align}
    \lambda_1 &= \frac{1}{2} \left( T_{11} + T_{22} + \sqrt{4T_{12}^2+(T_{11}-T_{22})^2} \right)\,,\label{eq:l1}\\
    \lambda_2 &= \frac{1}{2} \left( T_{11} + T_{22} - \sqrt{4T_{12}^2+(T_{11}-T_{22})^2} \right)\,.\label{eq:l2}
\end{align}
In the eigendecomposition, we write the deformation tensor as
\begin{align}
    \bm{\psi} = \begin{pmatrix}
        c & -s\\
        s & c
    \end{pmatrix}^T
    \begin{pmatrix}
        \lambda_1 & 0\\
        0 & \lambda_2
    \end{pmatrix}
    \begin{pmatrix}
        c & -s\\
        s & c
    \end{pmatrix},
\end{align}
with $c = \cos \theta$ and $s = \sin \theta$ where $\theta$ is the orientation of the eigensystem spanned by the eigenvectors $\bm{v}_1=(c,s)$ and $\bm{v}_2=(-s,c)$, giving us a relation between two parametrizations of the deformation tensor $(T_{11}, T_{12}, T_{22})$ and $(\lambda_1,\lambda_2,\theta)$. The PDF of the eigenvalue fields follows from the distribution of the second-order derivatives of the random field in a point.

The second order derivatives $\bm{Y}=(T_{11}, T_{12}, T_{22})$ are normally distributed random variables with vanishing mean, and the covariance matrix
\begin{align}
    M = \left \langle \Delta\bm{Y}^T\Delta\bm{Y}\right \rangle
    = \frac{\sigma_2^2}{8}\begin{pmatrix}
        3 & 0 & 1\\
        0 & 1 & 0\\
        1 & 0 & 3
    \end{pmatrix}.\label{eq:M1}
\end{align}
where the $\sigma_i$'s are defined for the power spectrum of the primordial displacement potential. The exponent of the normal distribution now takes the form
\begin{align}
    -\frac{1}{2} \Delta \bm{Y}^T M^{-1}\Delta\bm{Y} 
    &= -\frac{1}{2\sigma_2^2}\left( 3\left(T_{11} + T_{22}\right)^2 - 8 \left(T_{11}T_{22} - T_{12}^2\right) \right)\\
    &= -\frac{1}{2\sigma_2^2} \left( 3(\lambda_1 + \lambda_2)^2 - 8 \lambda_1 \lambda_2 \right),
\end{align}
which does not depend on the angle $\theta$ mirroring the statistical isotropy of the random field. As the Jacobian of the transformation from the linear coordinates $(T_{11},T_{12},T_{22})$ to the eigenframe coordinates $(\lambda_1,\lambda_2,\theta)$ takes the form $\mathrm{d}T_{11}\wedge\mathrm{d}T_{12}\wedge\mathrm{d}T_{22} = (\lambda_1-\lambda_2) \mathrm{d}\lambda_1 \wedge\mathrm{d}\lambda_2 \wedge \mathrm{d}\theta$, we obtain (after integrating over the angle $\theta$) the two-dimensional extension of the Doroshkevich formula
\begin{align}
    p(\lambda_1,\lambda_2) = \sqrt{\frac{2}{\pi}} \frac{2}{\sigma_2^3}|\lambda_1-\lambda_2| e^{-\frac{1}{2\sigma_2^2} \left( 3(\lambda_1 + \lambda_2)^2 - 8 \lambda_1 \lambda_2 \right)}\,.
\end{align}

The eigenvalue fields at a point are strongly correlated (see figure \ref{fig:Doroshkevich}). For example, one is unlikely to find the eigenvalue fields to coincide as this occurs with vanishing probability. The marginal distributions for the first eigenvalue field
\begin{align}
    p(\lambda_1) &=\frac{e^{-\frac{2 \lambda_1^2}{\sigma_2^2}}}{9\sigma_2^2}\left(
        \sqrt{\frac{72}{\pi}}  \sigma_2 + 4 \sqrt{3} \lambda_1 e^{\frac{2 \lambda_1^2}{3 \sigma_2^2}}\left( 1+\text{erf}\left[\sqrt{\frac{2}{3}}\frac{\lambda_1}{\sigma_2}\right]\right)
    \right)\,,
\end{align}
follows a bell curve centered at the mean $\bar{\lambda}_1 = \langle\lambda_1 \rangle =+\sqrt{\pi}\sigma_2/4$ (see figure \ref{fig:PDF}). The distribution of the second eigenvalue field 
\begin{align}
    p(\lambda_2) &=\frac{e^{-\frac{2 \lambda_2^2}{\sigma_2^2}}}{9\sigma_2^2}\left(
        \sqrt{\frac{72}{\pi}}  \sigma_2 - 4 \sqrt{3} \lambda_2 e^{\frac{2 \lambda_2^2}{3 \sigma_2^2}}\text{erfc}\left[\sqrt{\frac{2}{3}}\frac{\lambda_2}{\sigma_2}\right]
    \right)
\end{align}
is its mirror image with the mean $ \bar{\lambda}_2 = \langle\lambda_2 \rangle =-\sqrt{\pi}\sigma_2/4$ (see figure \ref{fig:PDF}). The PDF of the eigenvalue fields is tighter than the PDF of the corresponding density perturbation, as the sum of the eigenvalue fields coincides with the density perturbation (see equation \eqref{eq:lambda_delta}).

\begin{figure}
    \centering
    \includegraphics[width=\linewidth]{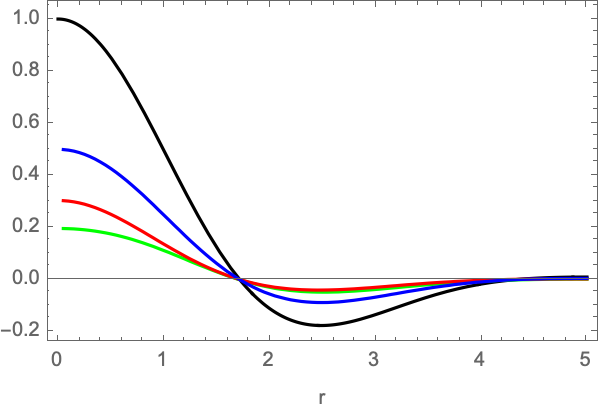}
    \caption{The two-point correlation functions of the density perturbation and eigenvalue fields. The two-point correlation function of the density perturbation $\xi_\delta$ (black), the autocorrelation functions of the eigenvalue fields (red), the crosscorrelation function of the two eigenvalue fields (green), and the crosscorrelation functions of the eigenvalue fields with the density perturbation.}\label{fig:lambda_correlations}
\end{figure}

\begin{figure*}
    \vskip -0.5cm
    \centering
    \begin{subfigure}[b]{0.49\textwidth}
        \includegraphics[width=0.9\linewidth]{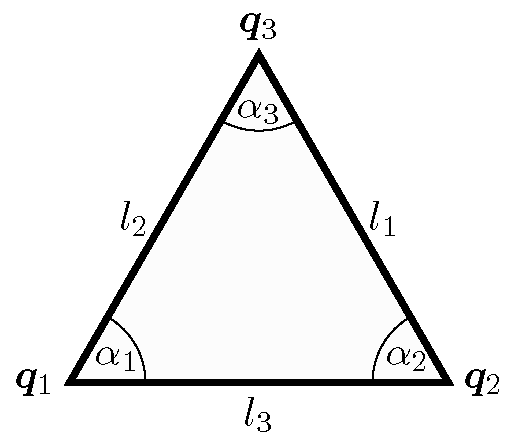}
        \caption{The configuration.}\label{fig:triangle}
    \end{subfigure}~
    \begin{subfigure}[b]{0.4\textwidth}
        \includegraphics[width=0.9\linewidth]{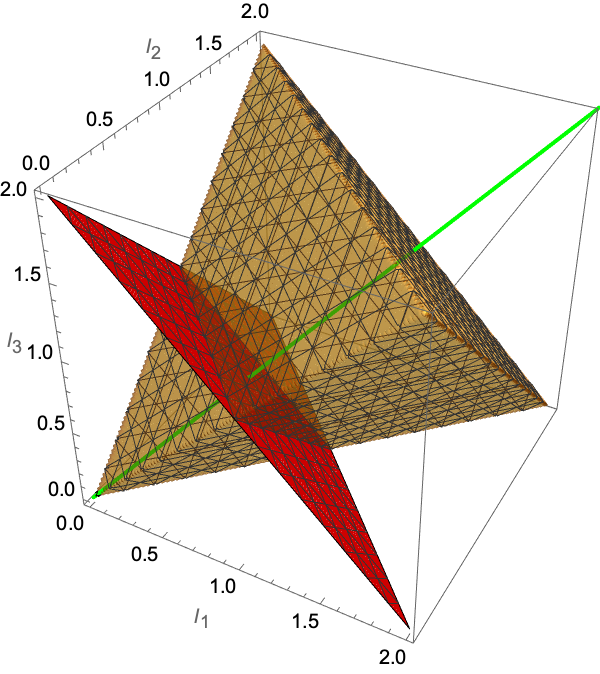}
        \caption{The triangle conditions in the space of side-lengths.}\label{fig:lll}
    \end{subfigure}
    \caption{Triangle configurations for the three-point correlation functions. The left panel shows the parametrization of a triangle in terms of the points, sidelengths, and angles. In the right panel, the yellow wedge consists of the side lengths $(l_1,l_2,l_3)$ corresponding to a triangle configuration. The red plane represents the space of constant circumference $l_T=l_1+l_2+l_3$. The green line corresponds to the equilateral triangle configurations.}
    \vskip 0.25cm
    \centering
    \includegraphics[width=0.6\textwidth]{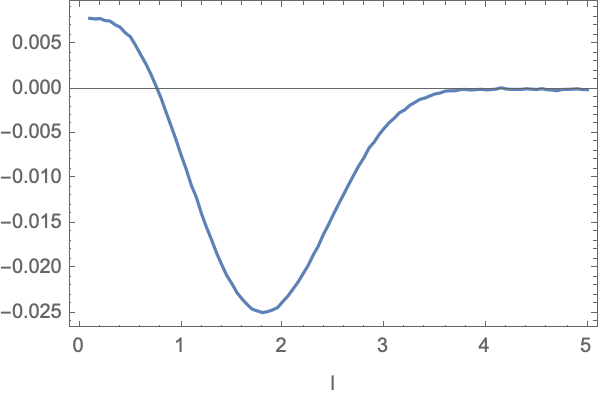}
    \caption{The three-point correlation function of the first eigenvalue field for equilateral triangle configurations $(l_1,l_2,l_3)=(l,l,l)$ in units of $R_s$.}\label{fig:equilateral}
    \vskip 0.25cm
    \centering
    \begin{subfigure}[b]{0.33\textwidth}
        \includegraphics[width=0.9\linewidth]{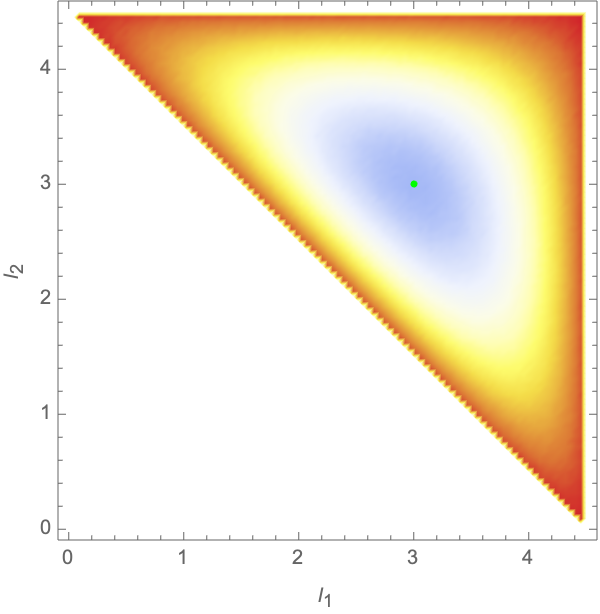}
        \caption{$l_T=3 \times 1$}
    \end{subfigure}~
    \begin{subfigure}[b]{0.33\textwidth}
        \includegraphics[width=0.9\linewidth]{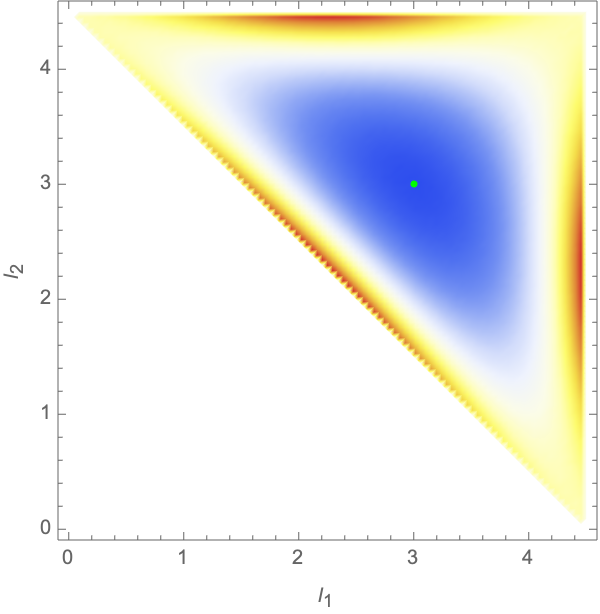}
        \caption{$l_T=3 \times 2$}
    \end{subfigure}~
    \begin{subfigure}[b]{0.33\textwidth}
        \includegraphics[width=0.9\linewidth]{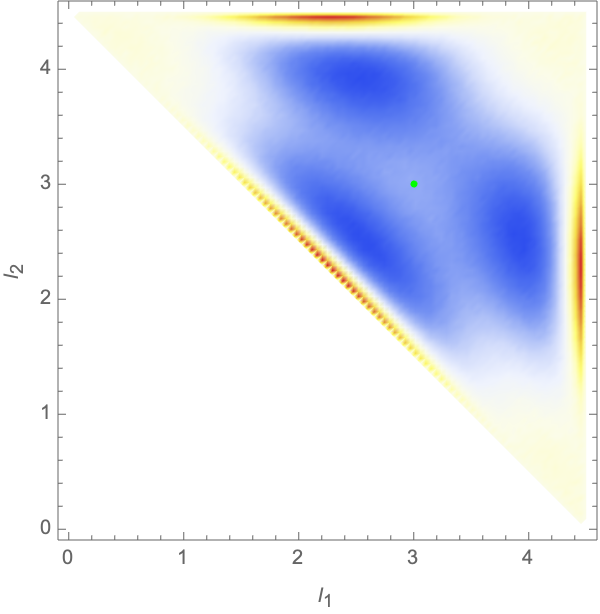}
        \caption{$l_T=3 \times 3$}
    \end{subfigure}~
    \caption{The three-point function of the first eigenvalue fields for the circumferences $l_T=3,6,9$. The centroid (the green point) represents the equilateral triangle configuration. The three corners are known as squeezed configurations. The three regions close to the midpoints of the edges are known as flattened triangle configurations.}\label{fig:threePoint}
\end{figure*}

\subsection{The two-point correlation functions}
The displacement potential is completely characterized by the two-point correlation function $\xi_\Psi$. The two-point correlation function of the eigenvalue fields, at two points $\bm{q}_1$ and $\bm{q}_2$ separated by $\bm{r}=\bm{q}_2-\bm{q}_1$, follows from the distribution of the second-order derivatives $T_{11}, T_{12}, T_{22}$ and $U_{11}, U_{12}, U_{22}$ at the two points. These variables are again normally distributed with a vanishing mean and the covariance matrix
\begin{align}
    M = \begin{pmatrix}
        N(\bm{0}) & N(\bm{r})\\
        N(\bm{r}) & N(\bm{0})
    \end{pmatrix}
\end{align}
with the $N(\bm{r})$ the covariance matrix of $(T_{11},T_{12},T_{22})$ and $(U_{11},U_{12},U_{22})$ given by
\begin{align}
    N(\bm{r}) = \begin{pmatrix}
        \partial_1^4 \xi_\Psi(\bm{r}) & 0 & \partial_1^2 \partial_2^2 \xi_\Psi(\bm{r})\\
        0 & \partial_1^2\partial_2^2\xi_\Psi(\bm{r}) & 0\\
        \partial_1^2\partial_2^2\xi_\Psi(\bm{r}) & 0 & \partial_2^4\xi_\Psi(\bm{r})
    \end{pmatrix}.\label{eq:N}
\end{align}
We explicitly derive the upper left element of this matrix. The correlation of the second-order derivatives $T_{11}$ and $U_{22}$ can be expressed as the fourth-order derivative of the two-point correlation function in the $x$-direction, \textit{i.e.}, 
\begin{align}
    \langle T_{11} U_{11} \rangle 
        &= \left \langle  \partial_1^2 f(\bm{q}_1) \partial_1^2 f(\bm{q}_2)\right \rangle\\
        &= \left\langle
        \int  k_{1}^2 e^{-i\bm{k}\cdot \bm{q}_1}\hat{\phi}(\bm{k})\frac{\mathrm{d}\bm{k}}{(2\pi)^2}
        \int  l_{1}^2 e^{i\bm{l}\cdot \bm{q}_2}\hat{\phi}^*(\bm{l})\frac{\mathrm{d}\bm{l}}{(2\pi)^2} \right\rangle\\
        &= \frac{1}{(2\pi)^4} \iint 
        k_1^2 l_1^2
        \left \langle \hat{\phi}(\bm{k})\hat{\phi}^*(\bm{l})\right \rangle e^{-i\bm{k}\cdot\bm{q}_1 + i \bm{l}\cdot \bm{q}_2}
        \mathrm{d}\bm{k}\mathrm{d}\bm{l}\\
        &=\frac{1}{(2\pi)^2} \int k_1^4 P_\Psi(\bm{k})
        e^{-i\bm{k}\cdot(\bm{q}_1-\bm{q}_2)}\mathrm{d}\bm{k}\\
        &=\partial_1^4 \xi_\Psi(\bm{q}_1-\bm{q}_2)\,,
\end{align}
where $\bm{k}=(k_1,k_2)$ and $\bm{l}=(l_1,l_2)$. The other components follow analogously. The covariance matrix is a generalization of the one at a single point. Indeed, in the limit $\bm{r} \to \bm{0}$, we recover equation \eqref{eq:lambda_delta} in terms of the generalized moment $\sigma_2$. For example, for the upper left component, we find $\partial_1^4 \xi_\Psi(\bm{r}) \to 3\sigma_2^2/8$ in the limit $\bm{r}\to\bm{0}$,
as
\begin{align}
    \lim_{\bm{r}\to \bm{0}} \partial_1^4 \xi_\Psi(\bm{q}_1-\bm{q}_2)
    &=\frac{1}{(2\pi)^2} \lim_{\bm{r}\to \bm{0}} \int k_1^4 P_\Psi(\bm{k}) e^{i\bm{k}\cdot\bm{r}}
        \mathrm{d}\bm{k}\\
    &=\frac{\sigma_2^2}{2\pi} \int_0^{2\pi} \cos^4\theta \mathrm{d}\theta \\
    &=\frac{3 \sigma_2^2}{8}
\end{align}

Using the definition of the eigenvalue fields in terms of the second-order derivatives (equations \eqref{eq:l1} and \eqref{eq:l2}), we obtain the autocorrelation function
\begin{align}
    &\left\langle (\lambda_i(\bm{q}_1) - \bar{\lambda}_i)(\lambda_j(\bm{q}_2) - \bar{\lambda}_j) \right\rangle\nonumber\\
    &= \int (\lambda_i(\bm{q}_1) - \bar{\lambda}_i)(\lambda_j(\bm{q}_2) - \bar{\lambda}_j)p(T_{11},T_{12},T_{22},U_{11},U_{12},U_{22})\nonumber \\
    &\phantom{=\int}\times \mathrm{d}T_{11}\mathrm{d}T_{12}\mathrm{d}T_{22}\mathrm{d}U_{11}\mathrm{d}U_{12}\mathrm{d}U_{22}\,.
\end{align}
The autocorrelation function of the two eigenvalue fields $\langle(\lambda_1(\bm{q}_1)-\bar{\lambda}_1)(\lambda_1(\bm{q}_2)-\bar{\lambda}_1)\rangle$ and $\langle(\lambda_2(\bm{q}_1)-\bar{\lambda}_2)(\lambda_2(\bm{q}_2)-\bar{\lambda}_2)\rangle$ coincide by the statistical symmetry of the eigenvalue fields (see the red curve in figure \ref{fig:lambda_correlations}). The autocorrelation function crosses zero at the same length scale as the two-point correlation function of the primordial density perturbation $\xi_\delta$. At the moment, we do not have a first principled explanation of this observation, in particular since the relevant correlation functions of the second-order derivatives of the displacement potential do not share this property. The crosscorrelation function of the eigenvalues $\langle(\lambda_1(\bm{q}_1)-\bar{\lambda}_1)(\lambda_2(\bm{q}_2)-\bar{\lambda}_2)\rangle$ and the density perturbations $\langle(\lambda_i(\bm{q}_1)-\bar{\lambda}_i)\delta(\bm{q}_2)\rangle$ are qualitatively similar to the cross-correlation functions (see the green and blue curve in figure \ref{fig:lambda_correlations}). Note that the cross-correlation function of the first and the second eigenvalue fields with the density perturbation coincide due to the statistical symmetry of the problem.

\subsection{The three-point correlation functions}
According to Isserlis-Wick's theorem, the correlation functions of a Gaussian random field can always be expressed in terms of the two-point correlation function. The odd correlation functions vanish and the higher-order even correlation functions coincide with a sum over products of the two-point correlation functions. For example, the three-point function vanishes,
\begin{align}
    \langle f(\bm{q}_1)f(\bm{q}_2)f(\bm{q}_3)\rangle = 0\,,
\end{align}
and the four-point function takes the form
\begin{align}
    \langle f(\bm{q}_1)f(\bm{q}_2)f(\bm{q}_3)f(\bm{q}_4)\rangle &= 
    \xi(\bm{q}_1-\bm{q}_2)\xi(\bm{q}_3-\bm{q}_4)\nonumber\\
    &\phantom{=}+
    \xi(\bm{q}_1-\bm{q}_3)\xi(\bm{q}_2-\bm{q}_4)\nonumber\\
    &\phantom{=}+
    \xi(\bm{q}_1-\bm{q}_4)\xi(\bm{q}_2-\bm{q}_3)\,.
\end{align}
As we saw in the previous section, the eigenvalue fields of the Hessian of a Gaussian random field are non-Gaussian. We here study the extent of the non-Gaussianity by evaluating the three-point correlation function
\begin{align}
    \langle (\lambda_i(\bm{q}_1)-\bar{\lambda}_i(\bm{q}_1)(\lambda_j(\bm{q}_2)-\bar{\lambda}_j(\bm{q}_2)(\lambda_k(\bm{q}_3)-\bar{\lambda}_k(\bm{q}_3)\rangle\,,\label{eq:three-point}
\end{align}
for $i,j,k=1,2$.

We extend the calculation presented in the previous section to evaluate the three-point correlation function. First consider the three sets of second-order derivatives $(T_{11}, T_{12}, T_{22})$, $(U_{11}, U_{12}, U_{22})$, and $(W_{11}, W_{12}, W_{22})$ corresponding the three points $\bm{q}_1,\bm{q}_2,\bm{q}_3$. The statistic $\bm{Y}=(T_{11}, T_{12}, T_{22}, U_{11}, U_{12}, U_{22}, V_{11}, V_{12}, V_{22})$ follows a multi-normal distribution with vanishing mean and the covariance matrix
\begin{align}
    M = \begin{pmatrix}
        N(\bm{0}) & N(\bm{q}_1- \bm{q}_2) & N(\bm{q}_1-\bm{q}_3)\\
        N(\bm{q}_1- \bm{q}_2) & N(\bm{0}) & N(\bm{q}_2- \bm{q}_3)\\
        N(\bm{q}_1- \bm{q}_3)& N(\bm{q}_1- \bm{q}_3) & N(\bm{0}) 
    \end{pmatrix},
\end{align}
where the matrix $N(\bm{r})$ is defined in equation \eqref{eq:N}. For a given configuration of points $\bm{q}_1,\bm{q}_2,\bm{q}_3$, we evaluate equation \eqref{eq:three-point} by sampling this distribution with a Monte Carlo scheme.

The three points $\bm{q}_1,\bm{q}_2,\bm{q}_3$ form a triangle which is, using the statistical isotropy and homogeneity of the gravitational potential, most conveniently expressed in terms of the side lengths $l_1=\|\bm{q}_2 - \bm{q}_3\|,l_2=\|\bm{q}_1-\bm{q}_3\|,l_3=\|\bm{q}_1-\bm{q}_2\|$ (see figure \ref{fig:triangle}), satisfying the triangle conditions
\begin{align}
|l_1-l_2| &< l_3 < l_1+l_2\,,\\
|l_1-l_3| &< l_2 < l_1+l_3\,,\\
|l_2-l_3| &< l_1 < l_2+l_3\,,
\end{align}
and the angles 
\begin{align}
\cos \alpha_1 &= \frac{l_2^2 + l_3^2 - l_1^2}{2 l_2 l_3}\,,\\
\cos \alpha_2 &= \frac{l_1^2 + l_3^2 - l_2^2}{2 l_1 l_3}\,,\\
\cos \alpha_3 &= \frac{l_1^2 + l_2^2 - l_3^2}{2 l_1 l_2}\,.
\end{align}
The triangle conditions define a wedge in the space of side lengths (see figure \ref{fig:lll}). Configurations with $l_1 \approx l_2 \approx l_3$ are known as equilateral configurations (see the green line in figure \ref{fig:lll}). When the triangle has two short and one long side, the triangle is called flattened. When the triangle has one short and two long sides, the triangle is in a squeezed configuration. The set of triangles with constant circumference $l_T=l_1+l_2+l_3$ form a plane in the $(l_1,l_2,l_3)$ space (see the red plane in figure \ref{fig:lll}). This plane intersects the triangle condition wedge in a triangle. We will display this triangle in the $(l_1,l_2)$-plane.

 \begin{figure*}
    \centering
    \begin{subfigure}[b]{0.24\textwidth}
        \includegraphics[width=\textwidth]{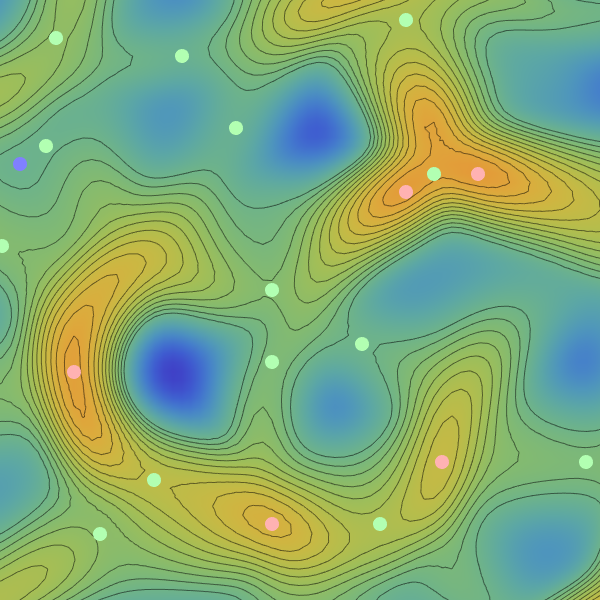}
    \end{subfigure}~
    \begin{subfigure}[b]{0.24\textwidth}
        \includegraphics[width=\textwidth]{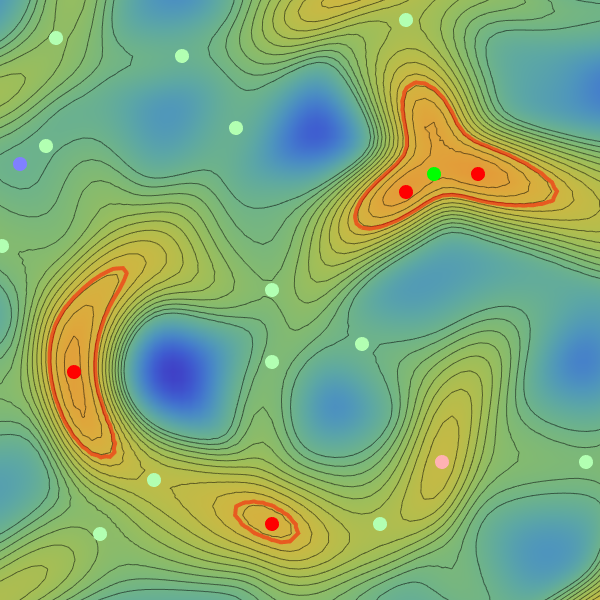}
    \end{subfigure}~
    \begin{subfigure}[b]{0.24\textwidth}
        \includegraphics[width=\textwidth]{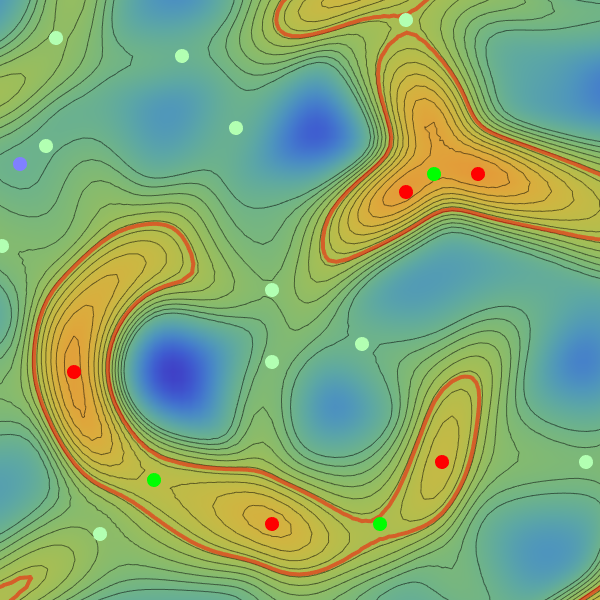}
    \end{subfigure}~
    \begin{subfigure}[b]{0.24\textwidth}
        \includegraphics[width=\textwidth]{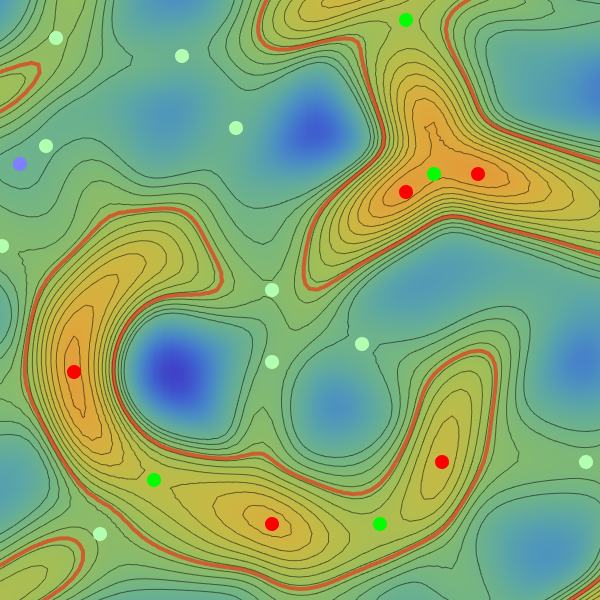}
    \end{subfigure}\\
    \begin{subfigure}[b]{0.24\textwidth}
        \includegraphics[width=\textwidth]{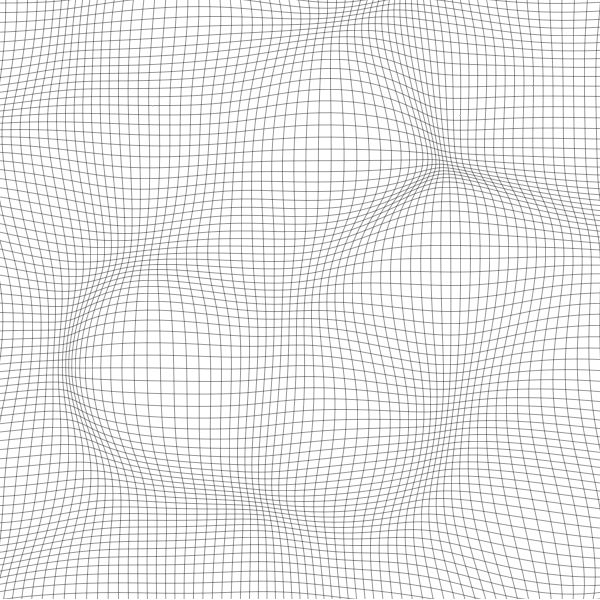}
    \end{subfigure}~
    \begin{subfigure}[b]{0.24\textwidth}
        \includegraphics[width=\textwidth]{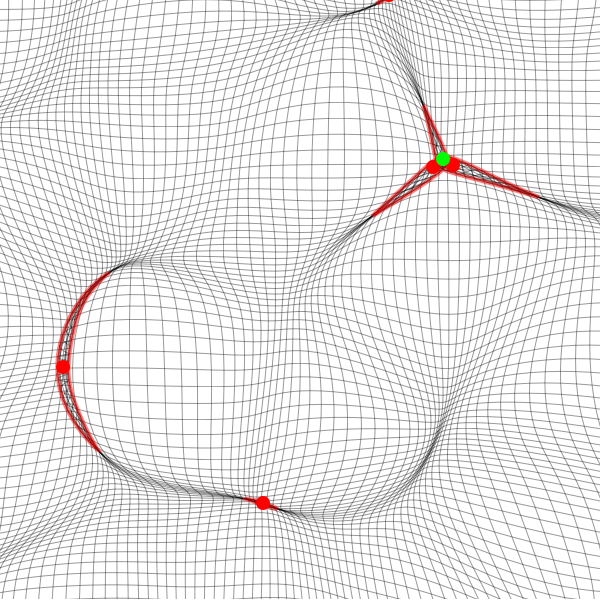}
    \end{subfigure}~
    \begin{subfigure}[b]{0.24\textwidth}
        \includegraphics[width=\textwidth]{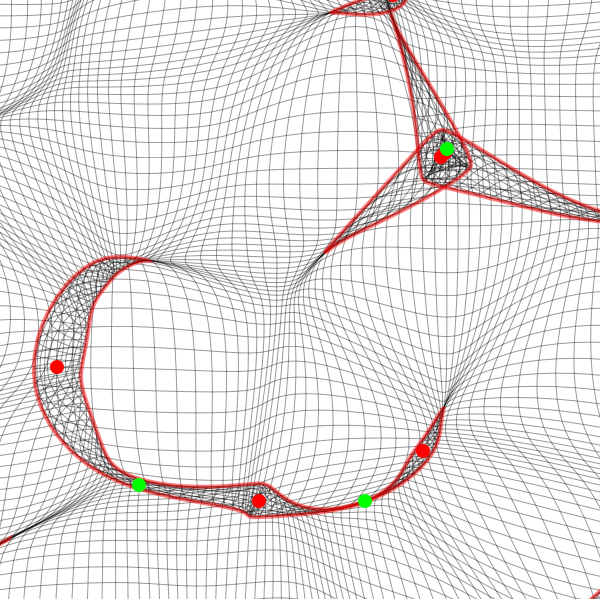}
    \end{subfigure}~
    \begin{subfigure}[b]{0.24\textwidth}
        \includegraphics[width=\textwidth]{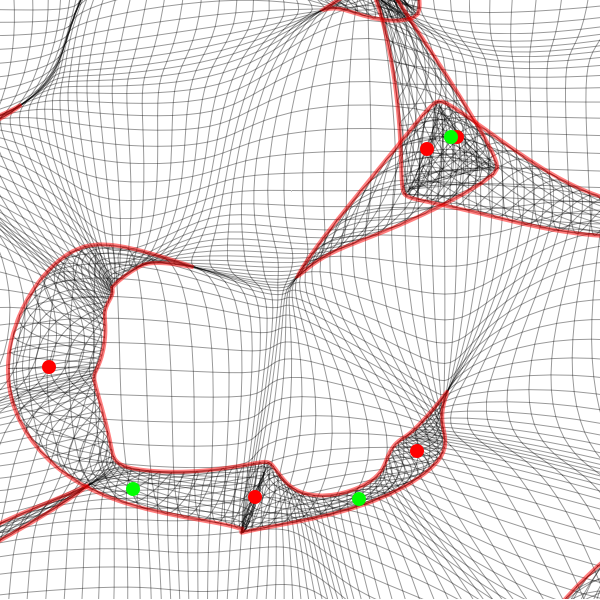}
    \end{subfigure}\\
    \begin{subfigure}[b]{0.24\textwidth}
        \includegraphics[width=\textwidth]{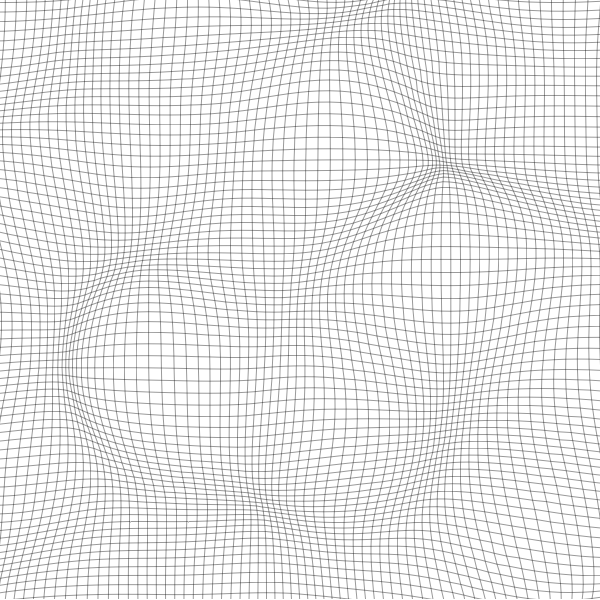}
    \end{subfigure}~
    \begin{subfigure}[b]{0.24\textwidth}
        \includegraphics[width=\textwidth]{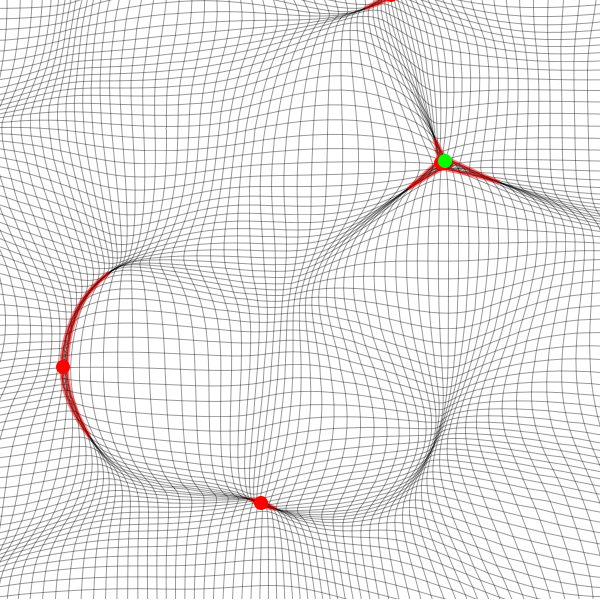}
    \end{subfigure}~
    \begin{subfigure}[b]{0.24\textwidth}
        \includegraphics[width=\textwidth]{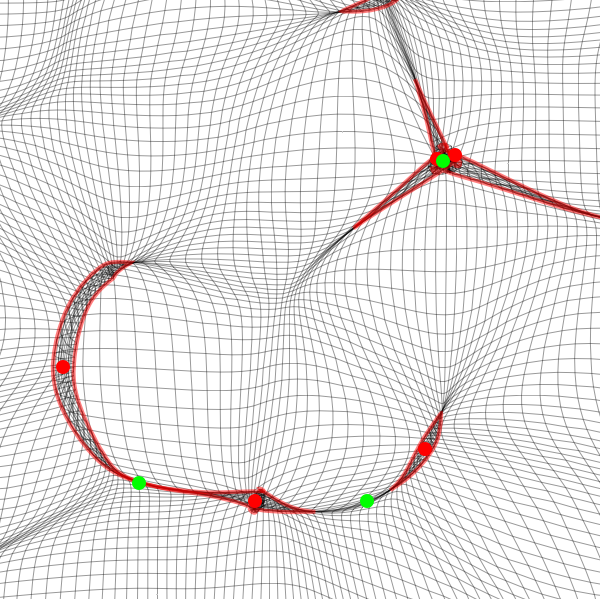}
    \end{subfigure}~
    \begin{subfigure}[b]{0.24\textwidth}
        \includegraphics[width=\textwidth]{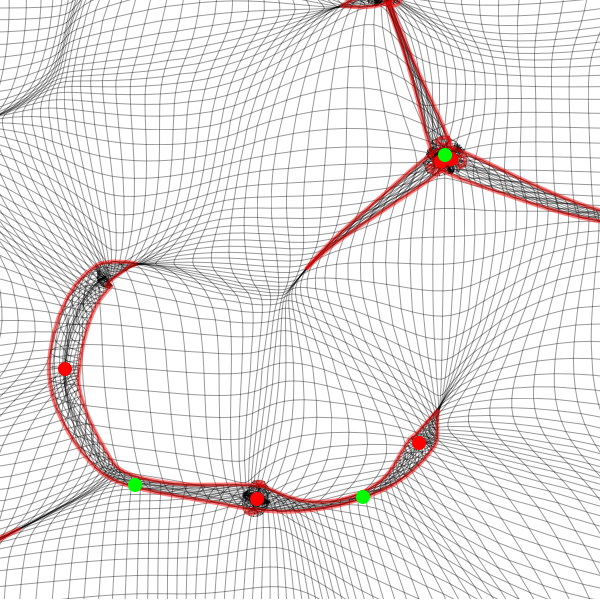}
    \end{subfigure}
    \caption{A two-dimensional simulation of the cosmic web at four stages of its evolution (left to right). The upper panels illustrate the first eigenvalue field with the corresponding singular points and critical curve corresponding to the fold caustic separating the single- from the multi-stream regions. The central panels illustrate the corresponding Zel'dovich approximation. The lower panels illustrate the corresponding non-linear universe in a $N$-body simulation.}\label{fig:cosmic_web}
\end{figure*}

The three-point correlation function of the first eigenvalue field for equilateral triangle configurations $(l_1,l_2,l_3)=(l,l,l)$ starts positive, crosses zero at a scale $l=0.8$, and is negative till about $l=3.5$ (see figure \ref{fig:equilateral}). At larger scales, the eigenvalue fields approach a Gaussian random field. Besides a strong contribution for equilateral configurations, we see that the three-point correlation function of the first eigenvalue field also peaks for flattened configurations (see figure \ref{fig:threePoint} for triangle configurations with constant circumference $l_T=3,6,9$). The three-point function does not have a strong squeezed component. At larger separations, the three-point correlation function vanishes signaling the scale at which the eigenvalue field is uncorrelated. This is in qualitative agreement with the vanishing two-point correlation function (see figure \ref{fig:lambda_correlations}).

\section{Singular points}\label{sec:singular}
The topology of the spatial cosmic matter distribution -- particularly its super-level set filtration -- is governed by its singular points. This is a fundamental insight from Morse theory \citep{Milnor:1963}, which states that the topology of a manifold changes at the function levels of the corresponding singular points. it is at these points that new topological features may emerge, features may merge or disappear.

For the primordial density perturbation, the singular points consist of its critical points which have over the last decades been extensively studied \citep{Bardeen:1986, Coles:1989, Baldauf:2016, Matsubara:2020a, Matsubara:2020b, Shim:2021, Rossi:2013}. The singularity structure of the primordial tidal and deformation tensor is more complex. It not only involves the critical points of the eigenvalue fields, as the singular points of the primordial tidal and deformation tensor include the cusp $A_3^\pm$ and umbilic $D_4^\pm$ points \citep[see][]{Arnold:1982a, Arnold:1982b, Hidding:2014, Feldbrugge:2018}. So far, these have received very little attention. In an assessment of the structure of the cosmic web based on its dynamics and dynamical evolution, we however have to address the statistical properties of these singularities. 

In this section, we first review the statistical properties of the spatial Gaussian density field. In line with the focus of this study, we subsequently extend the analysis to the distribution of the singular points of the primordial tidal and deformation tensor.

\subsection{Density field singular points}
The singular points of the primordial Gaussian density field comprise its critical points. A critical point is a point for which the gradient vanishes, \textit{i.e.},
\begin{align}
    \nabla \delta(\bm{q}_c) = \bm{0}\,.
\end{align}
The Hessian $\mathcal{H}\delta(\bm{q}_c)$ determines the nature of the point: the maxima, saddle points, and minima are defined as the critical points for which the Hessian has two, one, and relatively zero negative eigenvalues. The critical points of the density perturbations mark the change in the topology of the super-level set 
\begin{align}
    \delta^{-1}([\nu, \infty)) = \{\bm{q} \,|\, \delta(\bm{q}) \geq \nu\}.
\end{align}
The super-level set is empty for $\nu$ larger than the global maximum of the density perturbation. As $\nu$ is lowered, the local maxima lead to the introduction of disconnected components. The saddle points either lead to the merger of two disconnected components or the formation of a loop. The local minima fill the loops and remove them from the structure.

This process is neatly summarized by the Morse-Smale complex \citep{Morse:1925, Milnor:1963}, in which the maxima and minima are connected to the saddle points by integral lines (these satisfy steepest ascent and descent equations). These changes define alterations in the topology of the manifold. This may be related to the geometry of the mass distribution in terms of a natural skeleton of the cosmic web. The present-day cluster nodes, filaments, walls, and voids are associated with the maxima, two types of saddle points, and minima of the density field \citep[see for example][]{Aragon:2010, Sousbie:2011a, Sousbie:2011b}. More precisely, within Morse theory each of the critical points is associated with a manifold of steepest ascent and one of steepest descent, and the various constituent structures of the cosmic web are associated with these. 

\subsection{Singular points in tidal and deformation eigenvalue fields}
While the Morse-Smale complex of the density field may yield an intuitively suggestive characterization of the cosmic web at the corresponding cosmic epoch, a considerably more refined and dynamically relevant classification is the phase-space based classification of singularities of Caustic Skeleton theory \citep{Feldbrugge:2018}. It involves the singular points of the tidal/deformation field eigenvalues, and additional caustic singularities following from the
caustic conditions. 

\bigskip
\subsubsection{Caustic singularities}
Caustic skeleton theory yields a more refined classification of the cosmic web \citep{Arnold:1982a, Hidding:2014, Feldbrugge:2018}. It is a solid dynamical classification, that focuses on the formation of the different structural elements as a result of the folding and wrapping of the dark matter phase-space sheet in phase-space \citep{Shandarin:2011,Abel:2012,Falck:2012}. As we discussed in section \ref{sec:prelim}, the evolution of the cosmic web can be described in terms of the Lagrangian map
\begin{equation}
\bm{x}_t(\bm{q}) = \bm{q} + \bm{s}_t(\bm{q}). 
\end{equation}
The emergence of structural features around singularities is marked by the formation of multistream regions. This follows the appearance of regions where the density spikes as it collapses gravitationally (see equation \eqref{eq:densdens}). This goes along with the crossing of the corresponding mass streams, in a process called \textit{shell-crossing}, defined by
\begin{equation}
  \det \nabla \bm{x}_t =0\,.
\end{equation}
Following this, we see the emergence of multistream regions surrounded by \textit{Fold Caustics}. As the mass distribution evolves further, more complex features appear as a result of the merging of various multistream regions and the continuing wrapping into more complex multistream features of regions that had already collapsed.

The mathematics of \textit{catastrophe theory} allows the classification and identification of the complete set of caustic singularities that may emerge as a result of this process. It allows the classification of the different structural elements of the cosmic web on the basis of their dynamics and formation history. The singular points of the formation process are the points at which a multi-stream region is created, two such regions merge, or a multi-stream region disappears. In the two-dimensional case, there are two types of singular points, known as the \textit{Cusp} points and \textit{Umbilic} points. 

The simple analytical expressions of the Zeldovich approximation allow us to follow fully analytically the process of singularity formation and the resulting hierarchical buildup of structure and the embryonic skeleton of the cosmic web. This is enabled by the expression of the corresponding primordial tidal and deformation tensor $\bm{\psi}$. The implied density of an evolving mass element is the reciprocal of the determinant of the deformation tensor, $\det \nabla \bm{x}_t$,
\begin{equation}
  \delta(\bm{q}) \propto \frac{1}{\det \nabla \bm{x}_t}\,=\,\frac{1}{\det (I -b_+(t) \bm{\psi})}\,.
\end{equation}
The singular points consist of the cusp points defined as the critical points of the eigenvalue field $\lambda_i$ of the deformation tensor $\bm{\psi}$, and the umbilic points for which the two eigenvalue fields coincide $\lambda_1 = \lambda_2$. The topology of the multi-stream regions changes at a singular point when the eigenvalue in the singular point coincides with the reciprocal of the growing mode, \textit{i.e.}, 
\begin{align}
    \lambda_i = 1/b_+(t)    
\end{align}
Hence, the singular points relevant to the present cosmic web have an eigenvalue exceeding unity. Singular points which assume a negative eigenvalue are never realized, certainly not in the Zeldovich approximation. In summary, for following the dynamical evolution of the cosmic web, we need to address the eigenvalue singularity structure. 

\subsubsection{Caustic singularities and hierarchical structure formation}
The different roles of the primordial singular points of the tidal and deformation tensor can be visually inferred from figure \ref{fig:cosmic_web}. In addition to facilitating a dynamics and phase-space-based classification of the structural components of the cosmic web, it also implies a natural way of describing the hierarchical buildup of the cosmic web and understanding its connectivity characteristics: 
\begin{itemize}
    \item Each peak of the first eigenvalue field corresponds to the formation of a Zel'dovich pancake. In the upper panels of figure \ref{fig:cosmic_web}, we see the eigenvalue field with the corresponding local maxima (the red points). In the left column, the universe is still in a single-stream phase. At this stage, the Zeldovich approximation does an excellent job. We observe the onset of structure formation as the mass elements start to cluster. In the second column, two maxima in the lower left quadrant have undergone shell-crossing as the eigenvalue at these peaks exceeds $1/b_+(t)$. Each peak leads to the formation of a multi-stream region in the form of a Zeldovich pancake.
    \item Over time, these pancakes extend and merge at the saddle points of the first eigenvalue field (the green points). In transitioning from the second to the third column, we observe that two saddle points undergo shell-crossing connecting three Zeldovich pancakes. In general, the saddle points correspond to merger events of the multi-stream regions, forming the interconnected cosmic structure we observe in cosmological redshift surveys.
    \item At later times, we observe the breakdown of the Zeldovich approximation, as the configuration of the mass elements starts to diverge from the configuration in the $N$-body simulation. Even though, the Zeldovich approximation makes the multi-stream regions to puffy -- ignoring second infall -- it still does an accurate job at predicting the topology of the non-linear cosmic web. The local minima, or more explicitly the region of the first eigenvalue field below the threshold $1/b_+(t_0)$ have not undergone shell-crossing in the Zeldovich approximation at the current time. These regions can be associated with the voids of the cosmic web. When a valley assumes a value exceeding the level $1/b_+(t_0)$, the valley corresponds to a void that gets absorbed in a multi-stream structure.
    \item Finally, we draw attention to the triangular structure in the upper right quadrant of the figure. This is an example of an umbilic $D_4$ point. Shell-crossing takes simultaneously place along two directions. This is directly visible in the Zeldovich approximation by the occurrence of a triangular seven-stream region. In the $N$-body simulation, we see that the umbilic point corresponds to a node at which three filamentary structures meet.
\end{itemize}

\subsection{Stochastic geometry}
Before analyzing the spatial statistical properties of the singular points, we quickly review relevant theorems in stochastic geometry:
\begin{itemize}
    \item Rice's formula and its extensions \citep{Rice:1944, Rice:1945, Longuet-Higgins:1957, Adler:1981, Bardeen:1986, Adler:2009} expresses the number density of a set of isolated points, of a random field $f$ defined by the conditions $c_1[f](\bm{q}) =0, c_2[f](\bm{q})=0$, in terms of the expectation value 
    \begin{align}
        \mathcal{N} = \langle |\det \nabla \bm{c}[f] |\, \delta_D^{(2)}(\bm{c}[f])\rangle\,,
    \end{align}
    with the shorthand $\bm{c}[f]=(c_1[f],c_2[f])$. The determinant $|\det \nabla \bm{c}[f]|$ weights the different configurations satisfying the condition $\bm{c}[f]=0$. When we are interested in a refined set, satisfying an addition condition $d[f] >0$, Rice's formula assumes the form
    \begin{align}
        \mathcal{N} = \langle |\det \nabla \bm{c}[f] |\, \Theta(d[f])\delta_D^{(2)}(\bm{c}[f]) \rangle\,,
    \end{align}
    with the Heaviside theta function $\Theta$.
    \item The curve length density of a random field, defined as the length of the curve satisfying the condition $c[f](\bm{q})=0$ per unit area, can be expressed as the expectation value
    \begin{align}
        \mathcal{L}=\langle \|\nabla c[f]\|\, \delta_D^{(1)}(c[f])\rangle\,,
    \end{align}
    with the weighing $\|\nabla c[f]\|$ of the configurations satisfying the condition $c[f]=0$.
    \item The clustering of a random set of points is, to first-order, captured by the two-point correlation function. More precisely, the two-point correlation function $\xi_{1-2}$ is defined as the excess or deficit probability that one finds two relevant points separated by a distance $r$ with respect to an independently distributed set of points \citep{Peebles:1980, Coles:1989}, \textit{i.e.}, the probability $\mathrm{d}P$ of finding two points in two volume elements $\mathrm{d}V_1$ and $\mathrm{d}V_2$ separated by $r$ is given by
    \begin{align}
        \mathrm{d}P = n^2[1+\xi_{1-2}(r)]\mathrm{d}V_1 \mathrm{d}V_2\,,
    \end{align}
    with the number density of points $n$.
    
    Rice's formula can be extended to express the correlation function in terms of expectation values. Given two sets of isolated points $\mathcal{S}_1$, defined by the condition $\bm{c}_1[f]=0$, and $\mathcal{S}_2$, defined by two condition $\bm{c}_2[f]=0$, the two-point correlation function can be expressed as
    \begin{align}
        \xi_{1-2}(r) = \frac{\mathcal{N}_{1-2}(r)}{\mathcal{N}_1 \mathcal{N}_2} - 1\,,
    \end{align}
    with the numerator
    \begin{align}
        \mathcal{N}_{1-2}(r) &= \big\langle 
            |\det \nabla\bm{c}_1[f](\bm{q}_1)| \delta_D^{(2)}(\bm{c}_1[f](\bm{q}_1)) \nonumber\\
            &\phantom{=\big\langle} \times
            |\det \nabla\bm{c}_2[f](\bm{q}_2)| \delta_D^{(2)}(\bm{c}_2[f](\bm{q}_2)) \big\rangle\,,
    \end{align}
    where the first condition is applied to a point located at $\bm{q}_1$ and the second condition is applied to a point located at $\bm{q}_2$ with the separation $r=\|\bm{q}_1-\bm{q}_2\|$, and the denominator consisting of the two terms
    \begin{align}
        \mathcal{N}_1&=\left\langle |\det \nabla\bm{c}_1[f]| \delta_D^{(2)}(\bm{c}_1[f])\right\rangle\,, \\
        \mathcal{N}_2 &=\left\langle|\det \nabla\bm{c}_2[f]|\delta_D^{(2)}(\bm{c}_2[f]) \right\rangle\,.
    \end{align}
    
    The two-point correlation function $\xi_{1-2}$ is an extension of the covariance $\xi$ discussed in section \ref{sec:GRF} to point processes. When the two conditions $\bm{c}_1, \bm{c}_2$ separated by a distance $\bm{q}_1-\bm{q}_2$ are independent, the numerator factorizes and the two-point correlation function vanishes.
\end{itemize}

\subsection{Spatial statistics: density field singularities}
The number density of critical points of the density perturbation $\delta$ assuming a value $\nu$, defined by the conditions $c_i[f]=\partial_i f$, is given by the expectation value
\begin{align}
    \mathcal{N}_\delta(\nu) 
    &= \left\langle|\det\mathcal{H}\delta|\, \delta_D^{(1)}(\delta - \nu)\delta_D^{(2)}(\nabla \delta)\right\rangle\label{eq:density_crit}\\
    &= \int |\delta_{,11}\delta_{,22}-(\delta_{,12})^2|\, p_{\bm{Y}}(\nu,0,0,\delta_{,11},\delta_{,12},\delta_{,22})\nonumber\\
    &\phantom{=\int}\times \mathrm{d}\delta_{,11}\mathrm{d}\delta_{,12}\mathrm{d}\delta_{,22}\,,
\end{align}
with the linear statistic $\bm{Y}=(\delta,\delta_{,1},\delta_{,2},\delta_{,11},\delta_{,12},\delta_{,22})$, where the subscript denotes the partial derivative of the density perturbation at a point. The statistic follows a multi-normal distribution $p_{\bm{Y}}$ with vanishing mean and the covariance matrix
\begin{align}
    M_{\bm{Y}}
    =
    \begin{pmatrix}
    \sigma_2^2 & 0 & 0 & -\frac{\sigma_3^2}{2}  & 0 & -\frac{\sigma_3^2}{2}\\
    0 & \frac{\sigma_3^2}{2}  & 0 & 0 & 0 & 0\\
    0 & 0 & \frac{\sigma_3^2}{2}  & 0 & 0 & 0\\
    -\frac{\sigma_3^2}{2}  & 0 & 0 & \frac{3\sigma_4^2}{8} & 0 & \frac{\sigma_4^2}{8}\\
    0 & 0 & 0 & 0 & \frac{\sigma_4^2}{8}  & 0\\
    -\frac{\sigma_3^2}{2}  & 0 & 0 & \frac{\sigma_4^2}{8}  & 0 & \frac{3\sigma_4^2}{8} 
    \end{pmatrix}\,.
\end{align}
To evaluate this integral, we use the definition of conditional probabilities $p_{\bm{Y}}(\bm{Y}) = p_{\bm{Y}_1|\bm{Y}_2}(\bm{Y}_1|\bm{Y}_2) p_{\bm{Y}_2}(\bm{Y}_2)$, with the free statistic $\bm{Y}_1=(\delta_{,11},\delta_{,12},\delta_{,22})$ and the conditioned statistic $\bm{Y}_2=(\delta,\delta_{,1},\delta_{,2})$, to write
\begin{align}
    \mathcal{N}_\delta(\nu) &= p_{\bm{Y}_2}(\nu,0,0)\int|\delta_{,11}\delta_{,22}-(\delta_{,12})^2|\nonumber\\
    &\phantom{=} \times p_{\bm{Y}_1|\bm{Y}_2}(\delta_{,11},\delta_{,12},\delta_{,22}\,|\,\nu,0,0)\, \mathrm{d}\delta_{,11}\mathrm{d}\delta_{,12}\mathrm{d}\delta_{,22}\\
    &= p_{\bm{Y}_2}(\nu,0,0)\left\langle|\delta_{,11}\delta_{,22}-(\delta_{,12})^2|\,\big|\, \delta=\nu,\delta_i=0\right\rangle\,,\label{eq:trick}
\end{align}
which we evaluate using Monte Carlo integration. 

\begin{figure*}
    \centering
    \begin{subfigure}[b]{0.48\textwidth}
        \includegraphics[width=\textwidth]{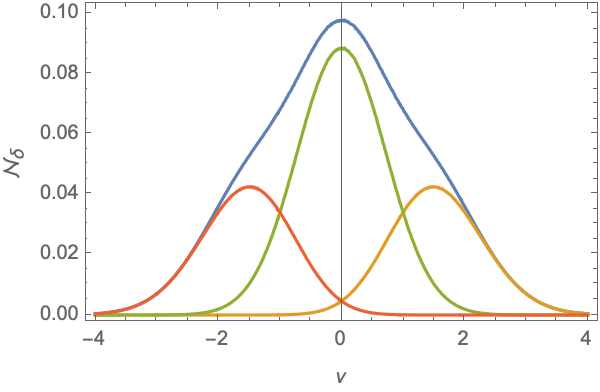}
        \caption{The number density of critical points of the primordial density perturbation $\delta$, with the peaks (orange), the saddle points (green), the valleys (red), and the collection of all critical points (blue).\\}\label{fig:critical_delta}
    \end{subfigure}~
    \begin{subfigure}[b]{0.48\textwidth}
        \includegraphics[width=\textwidth]{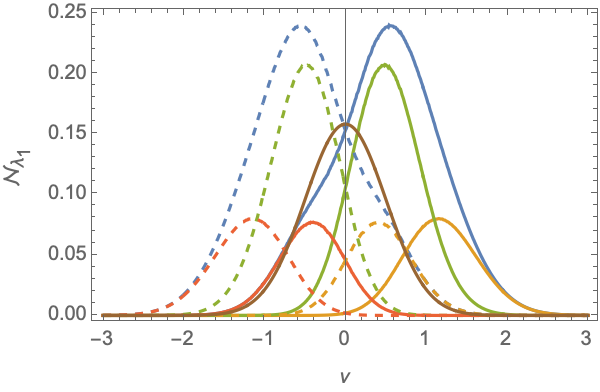}
        \caption{The number density of umbilic points (brown) and critical points of the first (solid) and second (dashed) eigenvalue fields $\lambda_i$, consisting of the peaks (orange), the saddle points (green), and the valleys (red). The total number density of critical points (blue) is the sum of the critical curves.}\label{fig:critical_D}
    \end{subfigure}
    \caption{The density of singular points of the primordial density perturbation (left) and the deformation tensor of the Zel'dovich approximation (right).}
\end{figure*}

We can further refine the number density to peaks, saddle points, and valleys by restricting the integration domain to configurations $(\delta_{,11},\delta_{,12},\delta_{,22})$ corresponding to a Hessian $\mathcal{H}\delta$ with two, one, or zero negative eigenvalues. Explicitly, we evaluate the expectation values
\begin{align}
    \mathcal{N}_{\delta,max}(\nu) 
    &= \left\langle|\det\mathcal{H}\delta|\, \delta_D^{(1)}(\delta - \nu)\delta_D^{(2)}(\nabla \delta)\Theta(\det \mathcal{H}\delta)\Theta(-\text{tr }\mathcal{H}\delta)\right\rangle\nonumber\,,\\
    \mathcal{N}_{\delta,sad}(\nu) 
    &= \left\langle|\det\mathcal{H}\delta|\, \delta_D^{(1)}(\delta - \nu)\delta_D^{(2)}(\nabla \delta)\Theta(-\det \mathcal{H}\delta)\right\rangle\nonumber\,,\\
    \mathcal{N}_{\delta,min}(\nu) 
    &= \left\langle|\det\mathcal{H}\delta|\, \delta_D^{(1)}(\delta - \nu)\delta_D^{(2)}(\nabla \delta)\Theta(\det \mathcal{H}\delta)\Theta(\text{tr }\mathcal{H}\delta)\right\rangle\nonumber\,,
\end{align}
for the number density of local maxima, saddle points, and local minima, using the fact that the determinant and trace of a two-by-two matrix coincide with the product and sum of the eigenvalues. See figure \ref{fig:critical_delta} for the resulting number densities of the primordial density perturbation. The number densities are close to Gaussian distributions. The number density of the saddle points is centered at $\nu=0$ and the number density of the local maxima is the mirror image of the number density of the local minima. 

\bigskip
The two-point correlation function of the critical points of the density perturbation takes the form
\begin{align}
    \xi_{\delta-\delta}(r;\nu_1,\nu_2) = \frac{\mathcal{N}_{\delta-\delta}(\nu_1,\nu_2)}{\mathcal{N}_\delta(\nu_1)\mathcal{N}_\delta(\nu_2)} -1\,,
\end{align}
the numerator is defined by the expectation value
\begin{align}
    \mathcal{N}_{\delta-\delta}(\nu_1,\nu_2)&=\langle \left|\det \mathcal{H}\delta_1 \right|\left|\det \mathcal{H}\delta_2 \right| \delta^{(1)}(\delta_1 - \nu_1) \delta^{(1)}(\delta_2 - \nu_2)\nonumber\\
    &\phantom{=\langle}\times \delta^{(2)}(\nabla_{\bm{q}} \delta_1) \delta^{(2)}(\nabla_{\bm{q}} \delta_2) \rangle
\end{align}
where $\delta_1$ and $\delta_2$ are the density perturbations at the two points $\bm{q}_1$ and $\bm{q}_2$ separated by a distance $r=\|\bm{q}_1-\bm{q}_2\|$, and the denominator is the product of two number densities of the relevant critical points discussed above.

\begin{figure}
    \centering
    \includegraphics[width=\linewidth]{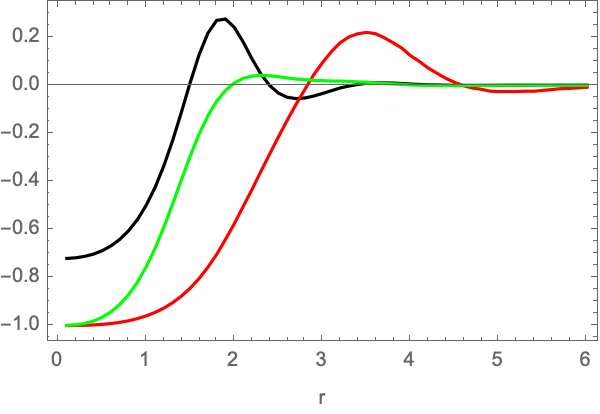}
    \caption{Two-point autocorrelation of critical points of the density perturbation $\delta$ in units of the smoothing scale $R_s$. The auto-correlation of all critical points (black), the peaks and valleys (red), and the saddle points (green).}\label{fig:twoPointDensity}
\end{figure}

See figure \ref{fig:twoPointDensity} for the resulting two-point correlation functions. We observe the characteristic correlation lengths of the critical points. For small separations, the probability of finding critical point pairs is smaller than what one would expect when the points are independently distributed. This is a consequence of the Gaussian smoothing of the density perturbation. For a separation $r=2R_s$, critical point pairs are more likely. For $r=3.5 R_s$, the two-point correlation function vanishes marking a scale at which the critical points are uncorrelated. When restricting the to local maxima/minima of the density perturbation, we see that the strongest correlation occurs for a slightly larger length scale. However, the qualitative behavior is similar.

\subsection{Spatial statistics: tidal and deformation singular points}
Next, we extend these results to the singular points of the deformation tensor.

\subsubsection{Eigenvalue field critical points}
The number density of critical points of the eigenvalue fields follows directly from Rice's formula
\begin{align}
    \mathcal{N}_{\lambda_i}(\nu) = \langle |\det \mathcal{H}\lambda_i|\, \delta^{(1)}(\lambda_i-\nu) \delta^{(2)}(\nabla \lambda_i)\rangle\,.
\end{align}
The main difference with respect to equation \eqref{eq:density_crit} is the non-linear nature of the eigenvalue fields and their derivatives. The eigenvalue fields can be defined in terms of the invariants of the deformation tensor,
\begin{align}
    T_{11} + T_{22} &= \lambda_1 + \lambda_2\,,\\
    T_{11}T_{22}-T_{12}^2&= \lambda_1 \lambda_2\,.
\end{align}
After differentiating these identities with the operators $\partial_1,$ $\partial_2,$ $\partial_1\partial_1,$ $\partial_1\partial_2,$ $\partial_2\partial_2$, we can solve for $\lambda_i, \partial_j \lambda_i,$ and $\partial_j \partial_k \lambda_i$ with $i,j,k=1,2$, to express the eigenvalue fields and their derivatives in terms of the partial derivatives of the displacement potential $\Psi$. This generalizes the well-known identities \eqref{eq:l1} and \eqref{eq:l2}. Unfortunately, these equations cannot be concisely expressed in these coordinates. 

The eigenvalue field and its derivatives are most easily related to the derivatives of the primordial gravitational potential in the eigenframe of the deformation tensor. In the derivation of the Doroshkevich formula, we used the eigenframe coordinates, mapping $(T_{11},T_{12},T_{22})$ to $(\lambda_1,\lambda_2,\theta)$. We here generalize this to the higher-order derivatives of the displacement potential. Given a smooth function $f$, a rotation $(q_1,q_2) \mapsto (\cos \theta q_1+ \sin \theta q_2, -\sin \theta q_1 + \cos \theta q_2)$ leads to a transformation of the partial derivatives, \textit{i.e.},
\begin{align}
    \partial_1 f(R \bm{q}) &= \cos\theta f^{(1)}(R \bm{q})- \sin\theta f^{(2)}(R \bm{q})\,,\\
    \partial_2 f(R \bm{q}) &= \sin\theta f^{(1)}(R \bm{q})+ \cos\theta  f^{(2)}(R \bm{q})\,,
\end{align}
with 
\begin{align}
    R = \begin{pmatrix}
        \cos \theta & \sin \theta\\ -\sin \theta & \cos \theta
        \end{pmatrix},
\end{align}
where $f^{(i)}$ is the partial derivative in the $q_i$ direction. The transformation of the higher-order derivatives follows by iterating this procedure. When identifying the angle $\theta$ with the orientation of the eigenvector $\bm{v}_1$, we construct a coordinate transformation from the space of partial derivatives of the displacement tensor $(T_{i_1\dots i_k})$ to the space of derivatives $(\theta,t_{i_1\dots i_k})$ with $t_{i_1\dots i_k}$ the $k$th-order derivative in the directions of the eigenvector fields $\bm{v}_1$ and $\bm{v}_2$. In effect, the angle $\theta$ replaces the second order derivative $T_{12}$ which vanishes in the eigenframe. The eigenvalues are expressed as the second-order derivatives of the displacement potential in the eigenframe, \textit{i.e.}, $\lambda_1=t_{11}$ and $\lambda_2=t_{22}$.

In the eigenframe coordinates, the gradient of the first eigenvalue field in the Cartesian coordinates takes the form
\begin{align}
    \nabla \lambda_1 = \left(t_{111} \cos \theta - t_{112} \sin \theta, t_{112}\cos \theta + t_{111} \sin \theta\right)\,.
\end{align}
The trace and the determinant of the Hessian of the first eigenvalue field simplify to
\begin{align}
    \text{tr } \mathcal{H} \lambda_1&= t_{1111}+t_{1122}+\frac{2\left(t_{112}^2+t_{122}^2\right)}{t_{11}-t_{22}}\,,\\
    \det \mathcal{H}\lambda_1 &= t_{1111}t_{1122} -t_{1112}^2 \nonumber \\
    &\phantom{=}+ \frac{2\left(t_{1111}t_{122}^2 -2 t_{1112}t_{112}t_{122}+ t_{1122}t_{112}^2 \right)}{t_{11}-t_{22}}\,.
\end{align}
The derivatives of the second eigenvalue field follow from the replacement $1 \leftrightarrow 2$. The Jacobian of the transformation from the Cartesian to the eigenframe coordinates $|t_{11}-t_{22}|$ is unchanged when including the higher-order derivatives.

Now, using the statistical isotropy of the Gaussian random field, we align the coordinate system with the eigenframe to write the number density of critical points of the first eigenvalue field as
\begin{align}
    &\mathcal{N}_{\lambda_1}(\nu) 
    =
    \int \left|T_{1111}\left(T_{1122}+\frac{2T_{122}^2}{\nu -T_{22}}\right) - T_{1112}^2 \right| \pi (\nu-T_{22})\Theta(\nu -T_{22})\nonumber\\
    &\times p_{\bm{Y}}(T_{11}=\nu, T_{12}=0, T_{22},T_{111}=0, T_{112}=0, T_{122},\nonumber\\
    &\phantom{\times p_{\bm{Y}}(\ }T_{1111},T_{1112},T_{1122})
    \mathrm{d}T_{22}\mathrm{d}T_{122}\mathrm{d}T_{1111}\mathrm{d}T_{1112}\mathrm{d}T_{1122}\,,
\end{align}
with the linear statistic $\bm{Y}=(T_{11}$, $T_{12}$, $T_{22}$, $T_{111}$, $T_{112}$, $T_{122}$, $T_{1111}$, $T_{1112}$, $T_{1122})$, following a multi-normal distribution with vanishing mean and the covariance matrix
{\tiny
\begin{align}
    M_{\bm{Y}}=
    \begin{pmatrix}
    \frac{3 \sigma_2^2}{8} & 0 & \frac{\sigma_2^2}{8} & 0 & 0 & 0 & -\frac{5 \sigma_3^2}{16} & 0 & -\frac{\sigma_3^2}{16}\\
    0 & \frac{\sigma_2^2}{8} & 0 & 0 & 0 & 0 & 0 & -\frac{\sigma_3^2}{16} & 0\\
    \frac{\sigma_2^2}{8} & 0 & \frac{3 \sigma_2^2}{8} & 0 & 0 & 0 & -\frac{\sigma_3^2}{16} & 0 & -\frac{\sigma_3^2}{16}\\
    0 & 0 & 0 & \frac{5 \sigma_3^2}{16} & 0 & \frac{\sigma_3^2}{16} & 0 & 0 & 0\\
    0 & 0 & 0 & 0 & \frac{\sigma_3^2}{16} & 0 & 0 & 0 & 0\\
    0 & 0 & 0 & \frac{\sigma_3^2}{16} & 0 & \frac{\sigma_3^2}{16} & 0 & 0 & 0\\
    -\frac{5 \sigma_3^2}{16} & 0 & -\frac{\sigma_3^2}{16} & 0 & 0 & 0 & \frac{35 \sigma_4^2}{128} & 0 & \frac{5 \sigma_4^2}{128}\\
    0 & -\frac{\sigma_3^2}{16} & 0 & 0 & 0 & 0 & 0 & \frac{5 \sigma_4^2}{128} & 0\\
    -\frac{\sigma_3^2}{16} & 0 & -\frac{\sigma_3^2}{16} & 0 & 0 & 0 & \frac{5 \sigma_4^2}{128} & 0 & \frac{3 \sigma_4^2}{128}
\end{pmatrix}\,.\nonumber
\end{align}}
We evaluate this integral with a Monte-Carlo scheme using the conditional probability trick explained in the equation \eqref{eq:trick}. We can refine the number density to peaks, saddle points, and valleys of the eigenvalue field by restricting the integration domain to regions for which the Hessian $\mathcal{H}\lambda_i$ as two, one, or zero negative eigenvalues. Note that for peaks, the trace is negative while the determinant is positive. For saddle points, the determinant is negative. For valleys, the trace and the determinant are positive. The number density of critical points of the second eigenvalue field follows analogously. 

Figure \ref{fig:critical_D} plots the number of densities evaluated with Monte Carlo integration. Note that the number densities of critical points of the eigenvalue fields are no longer centered at the threshold $\nu=0$. Moreover, the symmetry between the density of minima and maxima is broken. Finally, note that the number densities of the critical points of the eigenvalue fields are tighter than those of the density perturbation reflecting the relation $\delta = \lambda_1+\lambda_2$.

\bigskip
The two-point correlation function of the critical points of the eigenvalue fields, above a threshold $\nu$, follows an analogous equation,
\begin{align}
    \xi_{\lambda_i-\lambda_i}(r;\nu) = \frac{\mathcal{N}_{\lambda_i-\lambda_i}(\nu)}{\mathcal{N}_{\lambda_i}(r, \nu)^2} -1\,,
\end{align}
with
\begin{align}
    \mathcal{N}_{\lambda_i-\lambda_j}(r, \nu)&=
    \bigg\langle
    |\det \mathcal{H}\lambda_i||\det \mathcal{H}\mu_j|
    \delta^{(2)}(\nabla \lambda_i)\delta^{(2)}(\nabla \mu_i)\nonumber\\
    &\phantom{=\bigg \langle}\times
    \Theta(\lambda_i-\nu)\Theta(\mu_i-\nu)
    \bigg\rangle
\end{align}
where $\lambda_i$ and $\mu_i$ are the eigenvalues corresponding to the two points $\bm{q}_1, \bm{q}_2$ separated by the distance $r=\|\bm{q}_1-\bm{q}_2\|$, and 
\begin{align}
    \mathcal{N}_{\lambda_i}(\nu) = 
    \bigg\langle
    |\det \mathcal{H}\lambda_i|
    \delta^{(2)}(\nabla \lambda_i)
    \Theta(\lambda_i-\nu)
    \bigg\rangle\,,
\end{align}
and 
\begin{align}
    \mathcal{N}_{\lambda_i-\lambda_j}(r, \nu)&=
    \bigg\langle
    |\det \mathcal{H}\lambda_i|\delta^{(2)}(\nabla \lambda_i)\Theta(\lambda_i-\nu) \nonumber\\
    &\phantom{=}\times|\det \mathcal{H}\mu_j|\delta^{(2)}(\nabla \mu_i)\Theta(\mu_i-\nu)
    \bigg\rangle\,.
\end{align}
In the eigenframe coordinates $(\theta_t,t_{11},t_{22},t_{111},t_{112},\dots,t_{2222})$ at $\bm{q}_1$ and $(\theta_u,u_{11},u_{22},u_{111},u_{112},\dots,u_{2222})$ at $\bm{q}_2$, where the small letters represent the partial derivatives with respect to the eigenvector fields, we find the explicit identities
\begin{align}
    &\mathcal{N}_{\lambda_1-\lambda_1}(\nu)\nonumber\\
    &=
    \bigg\langle
    \left|t_{1111}t_{1122}-t_{1112}^2 + \frac{2 t_{1111}t_{122}^2}{\nu_1 - t_{22}}\right|\nonumber\\
    &\phantom{=\bigg\langle}\times
    \left|u_{1111}u_{1122}-u_{1112}^2 + \frac{2 u_{1111}u_{122}^2}{\nu_2 - u_{22}}\right|\nonumber\\
    &\phantom{=\bigg\langle}\times \delta^{(1)}(t_{111})\delta^{(1)}(t_{112})\delta^{(1)}(u_{111})\delta^{(1)}(u_{112})\nonumber\\
    &\phantom{=\bigg\langle}\times 
    \Theta(t_{11}-\nu)\Theta(u_{11}-\nu)\nonumber\\
    &\phantom{=\bigg\langle}\times |t_{11}-t_{22}|\Theta(t_{11}-t_{22})|u_{11}-u_{22}|\Theta(u_{11}-u_{22})
    \bigg\rangle\,.
\end{align}
for the numerator and
\begin{align}
    \mathcal{N}_{\lambda_1}(\nu)
    &=
    \bigg\langle
    \left|t_{1111}t_{1122}-t_{1112}^2 + \frac{2 t_{1111}t_{122}^2}{\nu_1 - t_{22}}\right|\delta^{(1)}(t_{111})\delta^{(1)}(t_{112})\nonumber\\
    &\phantom{=\bigg\langle}\times 
    \Theta(t_{11}-\nu)|t_{11}-t_{22}|\Theta(t_{11}-t_{22})
    \bigg\rangle\,,
\end{align}
for the denominator. The expressions for the second eigenvalue field follow analogously. 

To evaluate this expression with Monte Carlo methods, we evaluate the distribution of the second-, third- and fourth-order derivatives of the deformation potential at the two points, and extend the procedure explained in equation \eqref{eq:trick}:
\begin{enumerate}
    \item First, sample the space of second-order derivatives at the two points $Y_1=(T_{11}, T_{12}, T_{22}, U_{11}, U_{12}, U_{22})$.
    \item Next, find the eigenvalues $\lambda_1,\lambda_2,\mu_1,\mu_2$ and their corresponding eigenvectors, defining the angles $\theta_1$ and $\theta_2$.
    \item Given the angles $\theta_1$ and $\theta_2$ we define the eigenframes at the points $\bm{q}_1$ and $\bm{q}_2$, and construct the conditional joint distribution $Y_2=(t_{111},t_{112},u_{111},u_{112})$ conditioned on the second-order derivatives sampled in step (i). Note that $Y_2$ conditioned on $Y_1$ is multi-normal distributed even though the complete set $Y_1$ and $Y_2$ is non-Gaussian due to the non-linear rotation to the eigenframe. Integration with respect to $t_{111},t_{112},u_{111},u_{112}$ yields the conditional distribution evaluated at $t_{111}=t_{112}=u_{111}=u_{112}=0$ due to the Dirac delta functions.
    \item Evaluate the distribution of $Y_3=(t_{122},$ $t_{1111},$ $t_{1112},$ $t_{1122},$ $u_{122},$ $u_{1111},$ $u_{1112},$ $u_{1122})$ conditioned on $Y_1$ and $Y_2=0$. The statistic $Y_1$ conditioned on $Y_1$ and $Y_2$ is again multi-normal distributed, while the complete set is not. By sampling this distribution, we obtain a representative sample of the relevant distribution.
    \item Finally, evaluate the product of determinants of the Hessian of the eigenvalue fields and evaluate the mean to perform the integration.
\end{enumerate}
This procedure shows that the expectation value over $Q[Y_1, Y_3] \delta_D(Y_2)$ with the function $Q=|\det \mathcal{H}\lambda_i||\det \mathcal{H}\mu_j|$ can be expressed in terms of a nested expectation value weighted by the conditional density $p(Y_2=0|Y_1)$, as demonstrated by the identity
\begin{align}
    &\int Q[Y_1,Y_3] \delta_D(Y_2)p(Y_1,Y_2,Y_3) \mathrm{d}Y_1\mathrm{d}Y_2\mathrm{d}Y_3\\
    &= \int Q[Y_1,Y_3] \delta_D(Y_2) p(Y_3|Y_1,Y_2) p(Y_2|Y_1)p(Y_1) \mathrm{d}Y_1\mathrm{d}Y_2\mathrm{d}Y_3\\
    &= \int \left[ \int Q[Y_1,Y_3]  p(Y_3|Y_1,Y_2=0) \mathrm{d}Y_3\right] p(Y_2=0|Y_1)p(Y_1) \mathrm{d}Y_1\\
    &= \mathbb{E}[ p(Y_2=0|Y_1)\ \mathbb{E}[Q[Y_1,Y_3]|Y_1,Y_2=0] ]\,.
\end{align}

\begin{figure*}
    \centering
    \begin{subfigure}[b]{0.48\textwidth}
        \includegraphics[width=\textwidth]{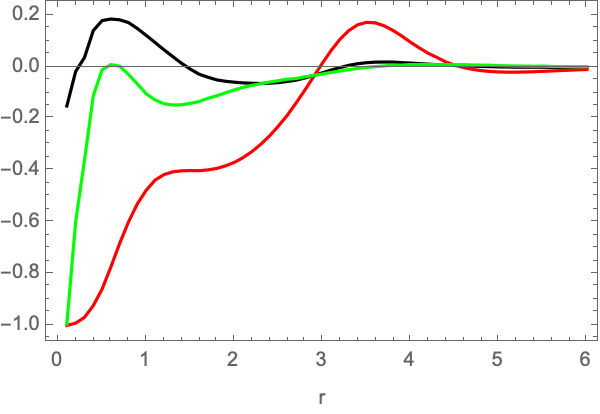}
        \caption{The two-point autocorrelation function of the singular points for $\nu=0$. All critical points (black), the local maxima (red), and saddle points (green).}\label{fig:singular1}
    \end{subfigure}
    \begin{subfigure}[b]{0.48\textwidth}
        \includegraphics[width=\textwidth]{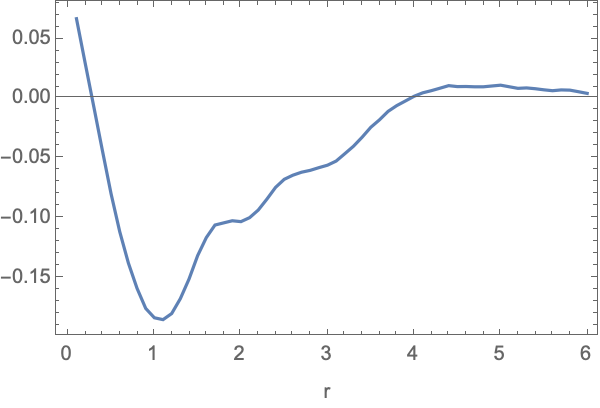}
        \caption{The two-point autocorrelation function of the singular points for $\nu=0$.\\$ $}\label{fig:D4D4_1}
    \end{subfigure}\\
    \begin{subfigure}[b]{0.48\textwidth}
        \includegraphics[width=\textwidth]{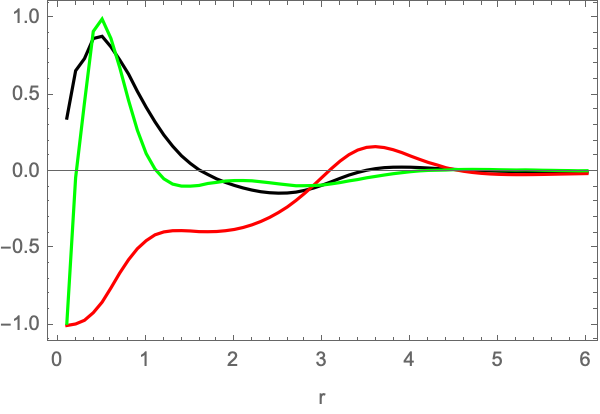}
        \caption{The two-point autocorrelation function of the singular points for $\nu=0.5$. All critical points (black), the local maxima (red), and saddle points (green).}\label{fig:singular2}
    \end{subfigure}
    \begin{subfigure}[b]{0.48\textwidth}
        \includegraphics[width=\textwidth]{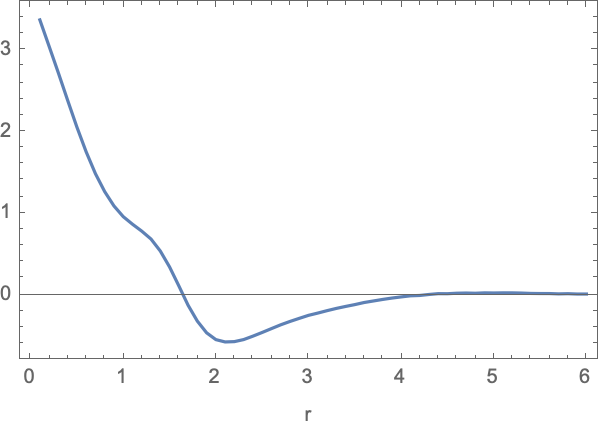}
        \caption{The two-point autocorrelation function of the singular points for $\nu=0.5$.\\$ $}\label{fig:D4D4_2}
    \end{subfigure}\\
    \begin{subfigure}[b]{0.48\textwidth}
        \includegraphics[width=\textwidth]{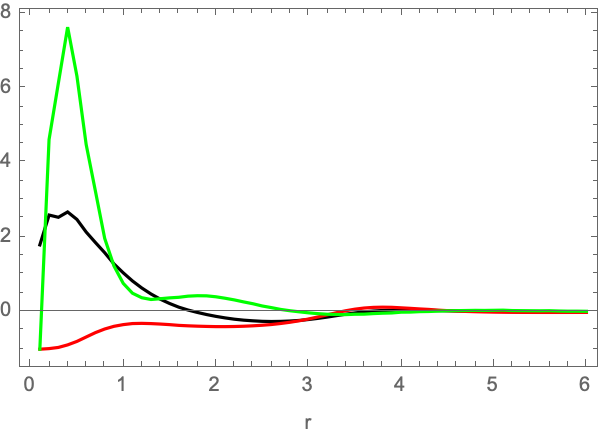}
        \caption{The two-point autocorrelation function of the singular points for $\nu=1$. All critical points (black), the local maxima (red), and saddle points (green).}\label{fig:singular3}
    \end{subfigure}
    \begin{subfigure}[b]{0.48\textwidth}
        \includegraphics[width=\textwidth]{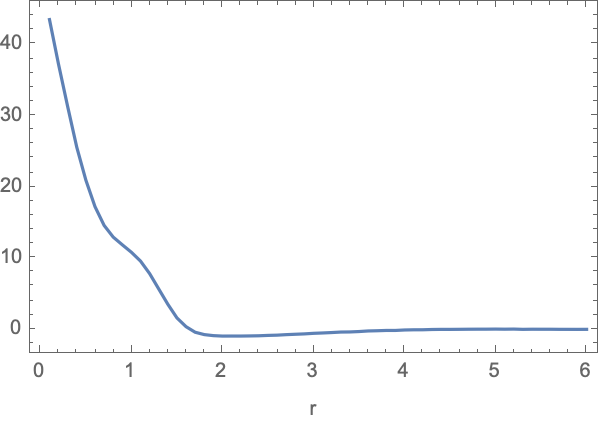}
        \caption{The two-point autocorrelation function of the singular points for $\nu=1$.\\$ $}\label{fig:D4D4_3}
    \end{subfigure}
    \caption{Two-point autocorrelation of singular points of the primordial deformation tensor, left the critical points and right the umbilic points, for the thresholds $\nu=0,0.5,$ and $1$.}
\end{figure*}

The resulting autocorrelation functions for three thresholds $\nu$ are plotted in figures \ref{fig:singular1}, \ref{fig:singular2}, and \ref{fig:singular3}. The correlation functions of the critical points of the eigenvalue fields are significantly richer than those of the critical points of the density perturbation, including a step-like structure from $r=R_s$ till about $r=2R_s$.

\bigskip
\subsubsection{Umbilic points}
The condition for the umbilic points, $\lambda_1=\lambda_2$, can be expressed in terms of two conditions on the partial derivatives of the primordial gravitational potential
\begin{align}
    T_{11}=T_{22}\,,\quad T_{12}=0\,
\end{align}
since the discriminant is the sum of two squares. That is to say, the eigenvalues 
\begin{align}
    \lambda_{1,2} = \frac{1}{2}\left(T_{11} + T_{22} \pm \sqrt{4T_{12}^2 + (T_{11}-T_{22})^2}\right)
\end{align} 
coincide if and only if the discriminant $4T_{12}^2 + (T_{11}-T_{22})^2$ vanishes. Using Rice's formula, we find the number density of umbilic points,
\begin{align}
    \mathcal{N}_{D_4}(\nu) 
    &= \langle \left|(T_{111}-T_{122})T_{122} - (T_{112}-T_{222})T_{112} \right| \nonumber\\
    &\phantom{=\langle}\times \delta^{(1)}(T_{11}-\nu)\delta^{(1)}(T_{11}-T_{22})\delta^{(1)}(T_{12})\rangle\\
    &= \int \left|(T_{111}-T_{122})T_{122} - (T_{112}-T_{222})T_{112} \right|\nonumber \\
    &\phantom{=\int}\times p_{\bm{Y}}(\nu,0,\nu,T_{111},T_{112},T_{122},T_{222})\nonumber\\
    &\phantom{=\int}\times\mathrm{d}T_{111}\mathrm{d}T_{112}\mathrm{d}T_{122}\mathrm{d}T_{222}\,,
\end{align}
with $p_{\bm{Y}}$ the density of the statistic $\bm{Y}=(T_{11},T_{12},T_{22},T_{111},T_{112},T_{122},T_{222})$. This yields the number density of umbilic points
\begin{align}
\mathcal{N}_{D_4}(\nu) 
&= 
\frac{2\sqrt{2}}{\pi^{3/2}\sigma_2^3} e^{-\frac{2 \nu^2}{\sigma_2^2}}
\int \left|(T_{111}-T_{122})T_{122} - (T_{112}-T_{222})T_{112} \right|\nonumber\\
&\phantom{=}\times p_{\bm{Y}_1}(T_{111},T_{112},T_{122},T_{222})\mathrm{d}T_{111}\mathrm{d}T_{112}\mathrm{d}T_{122}\mathrm{d}T_{222}\,.
\end{align}
See figure \ref{fig:critical_D} for the number densities of umbilic points. Note that only the singular points for which the eigenvalue field assumes a positive value can shell-crossing in the Zel'dovich approximation take place. 

\bigskip
The two-point correlation function of the umbilic points above a threshold $\nu$ is given by
\begin{align}
    \xi_{D_4-D_4}(r;\nu)=\frac{\mathcal{N}_{D_4-D_4}(r, \nu)}{\mathcal{N}_{D_4}(\nu)^2}-1\,,
\end{align}
where $\mathcal{N}_{D_4}(\nu)$ is the number density of umbilic points above the threshold $\nu$ and $\mathcal{N}_{D_4-D_4}(r, \nu)$ is the number density of pairs of umbilic points above the threshold separated by the distance $r$. Since the $D_4$ points can be defined by the conditions $T_{11}-T_{22}=0$ and $T_{12}=0$, Rice's formula yields the identities
\begin{align}
    \mathcal{N}_{D_4}(\nu) &= \big\langle 
        \left|
    (T_{111}-T_{122})T_{122}-(T_{112}-T_{222})T_{112}) \right|\nonumber\\
    & \phantom{=\big\langle}\times \delta_{D}^{(1)}(T_{12})\delta_D^{(1)}(T_{11}-T_{22})\Theta(T_{11}-\nu))
    \big\rangle\,,
\end{align}
and
\begin{align}
    \mathcal{N}_{D_4-D_4}(r, \nu) &= 
    \big\langle 
        \left|(T_{111}-T_{122})T_{122}-(T_{112}-T_{222})T_{112}) \right|\nonumber\\
        & \times \left|(U_{111}-U_{122})U_{122}-(U_{112}-U_{222})U_{112}) \right|\nonumber\\
        & \times \delta_{D}^{(1)}(T_{12})\delta_D^{(1)}(T_{11}-T_{22})\Theta( T_{11}- \nu))\nonumber\\
        & \times \delta_{D}^{(1)}(U_{12})\delta_D^{(1)}(U_{11}-U_{22})\Theta( U_{11}- \nu))
    \big\rangle\,.
\end{align}

The number density $\mathcal{N}_{D_4}(\nu)$ can be explicitly evaluated as the second- and third-order derivatives of the deformation potential are independently distributed. For the joined number density $\mathcal{N}_{{D_4}-{D_4}}$, evaluated the probability density for the Gaussian statistic $(T_{11},T_{12},T_{11}-T_{22},T_{111},T_{112},T_{122},T_{222},U_{11},U_{12},U_{11}-U_{22},U_{111},U_{112},U_{122},U_{222})$. After identifying the terms in the Dirac delta functions, we write this equation in terms of the density $p(T_{12}=0,T_{11}-T_{22}=0,\dots, U_{11}-U_{22}=0)$ and the conditional density $p(T_{11},T_{111},T_{112},\dots,U_{222}|T_{12}=0,T_{11}-T_{22}=0,\dots, U_{11}-U_{22}=0)$ over which we perform the Monte-Carlo sampling. Note that the evaluation of the two-point correlation function of the umbilic points is significantly simpler than the critical points of the eigenvalue fields, as their definitions are linear in the derivatives of the deformation potential.

The resulting autocorrelation functions for three thresholds $\nu$ are plotted in figures \ref{fig:D4D4_1}, \ref{fig:D4D4_2}, and \ref{fig:D4D4_3}. From the correlation functions, we see that umbilic points tend to cluster at small length scales. For the separation $r=2R_s$, we observe a deficit of umbilic pairs. For separations larger than $r=4R_s$, the umbilic points are nearly uncorrelated. For the threshold $\nu=0$, the two-point correlation function of the umbilic points consists of several plateaus. We do not observe this structure in the two-point correlation functions of the critical points of the primordial density perturbation. As the threshold $\nu$ is raised, these plateaus disappear. 
\section{Summary and discussion}\label{sec:Conclusion}
This study assesses the statistics of primordial (and Lagrangian) tidal and deformation tensor eigenvalue fields corresponding to Gaussian random density fields. It represents an extension of the well-known statistical characterization of Gaussian random fields \citep{Adler:1981, Bardeen:1986, Adler:2009} to that of intrinsically related non-Gaussian fields. The eigenvalue fields are distinctly non-Gaussian random fields, as evidenced by their well-known pdf derived by Doroshkevich \cite{Doroshkevich:1970}. In addition to the one-point functions of the eigenvalue values, we study the spatial structure in terms of two-point correlation functions and three-point correlation functions. The analysis concerns the statistical distribution for the continuous eigenvalue fields, as well as that of the singularities traced by the eigenvalue and eigenvector fields. 

The central incentive for studying the non-Gaussian statistics of tidal and deformation eigenvalue fields is the development of an analytical framework for analyzing the cosmic web, and to enable the exploitation of its complex geometric structure towards inferring information on cosmological parameters and the cosmic structure formation process. The cosmic web is a complex geometric interconnected pattern of voids, flattened walls, elongated filaments, and cluster nodes. It has formed through non-linear gravitational contraction and collapse of close to Gaussian initial conditions. Specific features of the primordial density field, such as its local maxima, minima, and saddle points, can be directly related to the different elements of the cosmic web. A variety of studies have attempted to describe its structure by investigating the properties of the density field and in particular the spatial distribution and connections of its singularity points \citep[see e.g]{Pogosyan:2009, Codis:2018, Shim:2021, Shim:2022}. While the latter provides insight into the weblike matter distribution at one cosmic epoch, it does not provide a path towards further insight into the gravitational evolution and hierarchical buildup, and connectivity, of the cosmic web and its various structural constituents. The present study is based on the realization that for a more encompassing description of the structure, connectivity, and hierarchical buildup of the cosmic web, we need to assess the more complex statistical properties of the tidal and deformation eigenvalue fields.  The observation that the spatial outline of the emerging cosmic web can already be recognized in the primordial tidal and deformation eigenvalue field (figure 1, \cite{Feldbrugge:2022}) forms a major justification for the analysis of its statistical properties presented in this study. 

The analysis of the tidal and corresponding deformation eigenvalue fields is crucial for our understanding of the formation and (hierarchical) evolution of the cosmic web. The instrumental role of tidal forces in shaping the cosmic mass distribution, and in particular in inducing the anisotropic gravitational contraction into its prominent filamentary and wall-like structures that delineates its spine, has been recognized by many studies \citep{Weygaert:1996, BondMyers:1996, Lee:1998, Catelan:2001, Porciani:2002a, Porciani:2002b, Hahn:2007, Weygaert:2008, Desjacques:2008, Rossi:2012, Paranjape:2018}. Insight into the spatial structure of the gravitational tidal force field induced by the inhomogeneous mass distribution, therefore, provides the physical basis for characterizing the spatial structure and pattern of the cosmic web.  Perhaps even more important is that it allows us to understand the dynamical evolution and buildup of the cosmic web, providing a solid physical basis for analyzing the spatial intricacies of the weblike matter distribution and its connectivity. It enables us to follow the temporal evolution of the cosmic web, and hence  supersedes the spatial characterization in terms of the density field and its singularities 

The analysis of the tidal eigenvalue fields also allows us to extend the dynamical description to a full phase-space-based inventory of the spatial patterns that arise in the evolving matter distribution. This is achieved by following the gravitational folding of the phase-space sheet, the mass distribution in the six-dimensional phase space \citep{Shandarin:2011, Abel:2012}, and identifying the singularity structure of its projection on Eulerian space. This rich singularity structure is in fact the cosmic web that we observe, with the various singularities corresponding to the various structural elements that make up the cosmic web. The development of the phase-space singularity structure follows directly from the tidally induced deformation of the mass elements, and by tracing of the multistream nature of the flow field that is the result of the emergence of these \textit{Caustic} singularities. To a considerable extent, one may follow this process accurately through the first-order Lagrangian perturbation theory, the Zeldovich approximation. In recent work, two of the authors developed the \textit{Caustic Skeleton} formalism to trace the multistream nature of the cosmic web by identifying the caustic singularities \citep{Feldbrugge:2018, Feldbrugge:2022}, building on early seminal work by Arnold, Zeldovich, and collaborators \citep{Arnold:1982a, Arnold:1982b}. The identity, location, and connectivity of the emerging caustic features follow from the geometric properties of the combination of \textit{eigenvalue} and \textit{eigenvector} fields of the primordial deformation tensor. The filamentary nature of the present-day cosmic web can therefore be traced back to the non-Gaussian nature of the eigenvalue field, and key dynamical events in large-scale structure formation  -- such as the creation and merger of multi-stream regions -- can be traced back to specific points in the initial conditions defined in terms of the primordial eigenvalue and eigenvector fields. 

In the present study, we analyze the properties of the primordial deformation tensor in a two-dimensional model of cosmic structure formation\footnote{The Caustic Skeleton model formalism is generically valid for any dimensional space, see Feldbrugge et al. 2018}. We extend the Doroshkevich formula to two-dimensional random fields and compare the correlation function of the eigenvalue fields with the correlation function of the density perturbation. For the spatial structure of the eigenvalue fields we first address their two-point correlation function. It is qualitatively similar to the two-point correlation function of the density perturbation. Their impact on the nontrivial spatial pattern of the cosmic web finds an expression in the corresponding three-point correlation function. The non-Gaussian nature of the eigenvalue field implies a non-zero three-point correlation function, showing both an equilateral and
a flattened component.

The study of the singularity structure related to the tidal and deformation eigenvalue field is significantly richer than that of the density field. It not only involves the maxima, minima, and saddle points of these fields, but also the singularity points of the implied caustic skeleton \citep{Feldbrugge:2018}. Most telling in this respect are the so-called umbilic points, points where two eigenvalues are equal. We analyze the number density of all classes of singular points, ie. of the critical points of the eigenvalue fields as well as the umbilic points. Following this, to describe their spatial clustering we here infer their two-point correlation function. The correlation functions show a richer structure than the correlation functions of the critical points in the primordial density perturbation, reflecting the non-Gaussian nature of the eigenvalue fields. We expect that these clustering properties refer to the geometry of the cosmic web.

In a future study, we will extend this study to three-dimensional Gaussian random fields and analyze the relation of the non-Gaussian nature of the eigenvalue fields to the present-day cosmic web. Moreover, we plan to go beyond the two-point correlation function and analyze how the evolving topology of the multi-stream regions in the cosmic web can be traced back to the eigenvalue fields and corresponding primordial caustic skeletons. We expect that the non-Gaussian nature of the eigenvalue fields may well have a stronger imprint on the topology than the two- and three-point correlation functions. It is through the intricate geometric and connectivity properties implied by the non-Gaussian tidal and deformation eigenvalue and eigenvector fields, explicitly expressed in the corresponding caustic singularity features, that we expect that the topological description in terms of Betti numbers and persistence diagrams may lead to new insights and new instruments for the study of the cosmic matter distribution \citep[see e.g.][]{Weygaert:2011, Pranav:2015, Wilding:2021, Wilding:2022}.

\section*{Acknowledgements}
JF is supported by the STFC Consolidated Grant ‘Particle Physics at the Higgs Centre’ and in part by the Higgs Fellowship.

\section*{Data Availability}
No new data were generated or analyzed in support of this research.



\bibliographystyle{mnras}
\bibliography{library} 



\appendix


\bsp	
\label{lastpage}
\end{document}